\documentclass[epj]{webofc}
\usepackage[varg]{txfonts}   
\usepackage{graphicx}
\usepackage{color}
\newcommand{\abs}[1]{\left| #1\right|}

\newcommand{\fnd}[2]{\frac{\textstyle #1}{\textstyle #2}}
\newcommand{\Imag}[1]{\Im {\it m}(#1 )}
\newcommand{\bm}[1]{\mbox{\boldmath $#1$}}
\woctitle{The 3rd International Conference on New Frontiers in Physics}
\begin{document}
\title{On the existence of a superlight scalar boson}
%
%

\author{Eef van Beveren\inst{1}\fnsep\thanks{\email{eef@teor.fis.uc.pt}}
\and
Susana Coito\inst{2}\fnsep\thanks{Present address:
Institute of Modern Physics, CAS, Lanzhou 730000, China;
\email{susana@impcas.ac.cn}}
\and
George Rupp\inst{2}\fnsep\thanks{\email{george@ist.utl.pt}}
}

\institute{Centro de F\'{\i}sica Computacional,
Departamento de F\'{\i}sica, Universidade de Coimbra,
P-3004-516 Coimbra, Portugal
\and
Centro de F\'{\i}sica das Interac\c{c}\~{o}es Fundamentais,
Instituto Superior T\'{e}cnico,
Universidade de Lisboa,
P-1049-001 Lisboa, Portugal
}

\abstract{%
In this lecture we show that the study of hadronic resonances
is severely hampered by the lack of accurate data
and, moreover, that for similar reason
the study of Weak substructure does not make sufficient progress.
We furthermore report on an unexplained high statistics signal
that may indicate the existence of a superlight scalar boson.
}
\maketitle
\section{Introduction}
\label{intro}

Half a century since the quark model was introduced
by Zweig \cite{CERNREPTH401/412} and Gell-Mann \cite{PL8p214},
a decade later it was followed by the Quantum Chromodynamics (QCD) proposal
of Fritzsch, Gell-Mann and Leutwyler \cite{PLB47p365}.
Several discoveries led to a deeper understanding of hadronic
interactions and their building blocks
\cite{PRL33p1404,PRL33p1406,PRL39p526,PRL40p671,
PRD21p2716,PLB76p361,PRL36p700,PRL39p252,PLB116p383}.
Nevertheless, we are still in the dark on the structure
of hadronic spectra, which are supposed to be a consequence
of the interactions between quarks/antiquarks and glue.
We basically live in an era of a plethora of suggestions
for the interpretation of hadronic scattering and/or production data.
But, as we will discuss in this work, the lack of accurate data
does not leave us sufficient hints for a clear direction
within the proliferation of models for mesonic and baryonic resonances.

Lattice QCD (LQCD) \cite{PRD10p2445},
which was suggested by Wilson shortly after QCD had been proposed
and since then has been explored by many researchers,
seemed to be the most powerful tool to link theory with experiment.
However, four decades later, it is recognised that its application
to hadronic production data still needs a much higher degree of perfection
than already achieved, leaving enough room for more modest strategies.

In the present work we discuss one of those latter strategies,
namely the Resonance-Spectrum Expansion (RSE) \cite{IJTPGTNO11p179}
for the description of mesonic spectra.
The RSE facilitates non-perturbative calculus
for scattering and production cross sections
and, furthermore, to find resonance pole positions
of the scattering amplitude in the complex invariant-mass plane.
Here, we will restrict ourselves to the Harmonic-Oscillator version
of the RSE (HORSE), since interhadronic dynamics,
governed by the glue, is, as we will see in the following,
well represented by harmonic-oscillator confinement
whereby the oscillator frequency $\omega$ is taken
equal to $\omega =190$ MeV,
independent of the meson's flavour content \cite{PRD21p772}.
It is assumed in the RSE that mesons couple to pairs of mesons
and vice versa by the creation/annihilation of $q\bar{q}$ pairs.
The intensity of pair creation/annihilation
is in RSE represented by a parameter $\lambda$
which in principle has to be adjusted to experiment,
but which in practice has been found to be rather independent
of the meson's flavour content \cite{PRD27p1527}.

\begin{figure}[htbp]
\begin{center}
\begin{tabular}{c}
\scalebox{0.8}{\includegraphics{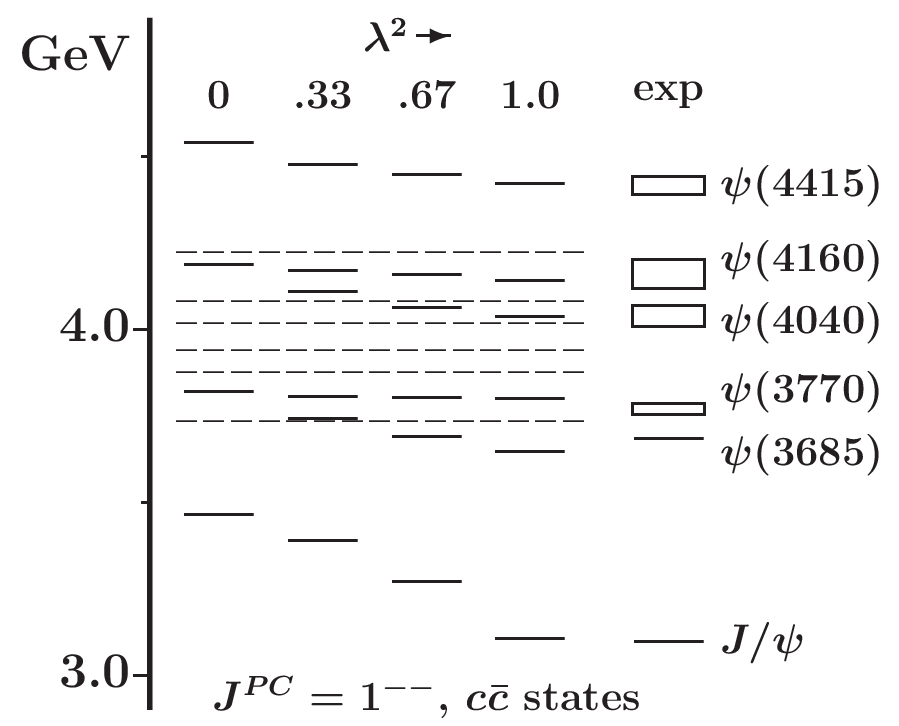}}
\end{tabular}
\end{center}
\caption[]{
The theoretical values \cite{PRD21p772} of the central resonance positions
for $J^{PC}=1^{--}$ charmonium $S$ and $D$ states
for various values of the model parameter
$\lambda$, compared to the experimental situation
\cite{PRL33p1404,PRL33p1406,PRL39p526,PRL40p671,
PRD21p2716,PLB76p361,PRL36p700}.
The short-dashed lines indicate the threshold positions
of the open--charm decay channels, respectively,
from the lowest to the highest threshold,
DD, DD$^{*}$, D$_{s}$D$_{s}$, D$^{*}$D$^{*}$,
D$_{s}$D$_{s}^{*}$ and D$_{s}^{*}$D$_{s}^{*}$.
The long dashes below $\lambda^{2}=0$ represent the masses
of the $c\bar{c}$ harmonic-oscillator spectrum, which are degenerate
for $S$ and $D$ states
except for the ground state at 3.47 GeV.
The long dashes below non-zero values of $\lambda^{2}$
indicate the real parts of the resonance poles.
The latter do not necessarily coincide with central
resonance positions for resonances above the DD threshold.
Note that the model treats on the same footing
bound states below and resonances above the DD threshold.}
\label{charm}
\end{figure}

In Fig.~\ref{charm} we show the results of HORSE
for the charmonium vector states.
One observes that the system of interacting
quarkonium and meson-meson channels
explains well the mass splittings
between charmonium $S$ and $D$ states.
It implies, moreover, that pure $S$ and $D$ states
do not exist in nature.
We verified that the dominantly $S$ states
end up at lower masses than the dominantly $D$ states.
In fact, the dominantly $D$ states are such admixtures
of $S$ and $D$ states that they almost decouple completely
from the meson-meson decay channels,
hence form narrow resonances which,
due to small mass shifts, do not deviate much
from the $c\bar{c}$ harmonic-oscillator spectrum.

It is well known that com\-po\-si\-te\-ness may be studied
from the appearance of resonance en\-han\-ce\-ments
in the event distributions of scattering and production experiments.
An extensive study on ha\-dro\-nic com\-po\-si\-te\-ness
published by Godfrey and Isgur in Ref.~\cite{PRD32p189}
gave us a good insight into the spectrum of quarkonia
obtained by the scattering of mesons
and by the event distributions of two or more hadrons
produced in production experiments.

Less well known is that the phenomenon of dynamically generated resonances,
like the light scalar mesons
$f_{0}(500)$, $K^{\ast}_{0}(800)$, $f_{0}(980)$ and $a_{0}(980)$,
is related to the main source for the coupling
of quark-antiquark systems to systems of two or more hadrons,
namely quark-pair creation and annihilation,
hence a consequence of ha\-dro\-nic com\-po\-si\-te\-ness.
Dynamically generated resonances do not make part
of the quarkonia spectrum.
Different configurations,
like multi-quark states \cite{PRD15p267}
or meson molecules \cite{PRD41p2236},
have been proposed.
Such configurations, however, form an integral part
of a complete description of scattering and production processes
based on the coupling, through quark-pair creation,
of quarkonia to multi-quark and multi-hadron systems.
It was found that the main components
which are necessary for a good description
of mesonic scattering and production data
are quarkonia and two-meson channels \cite{ZPC30p615}.

\begin{figure}[htbp]
\begin{center}
\begin{tabular}{c}
\scalebox{0.56}{\includegraphics{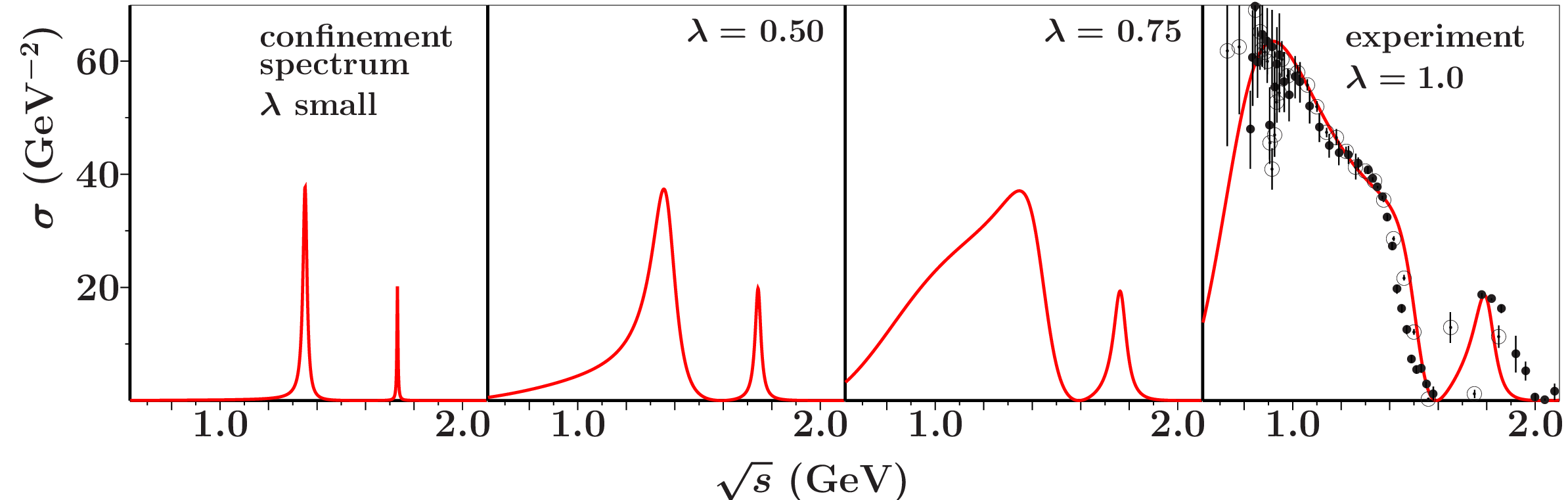}}
\end{tabular}
\end{center}
\caption[]{Theoretical \cite{ZPC30p615}
and experimental $S$-wave $K\pi$ scattering
cross sections in the isospin $I=1/2$ channel for different values
of the overall coupling $\lambda$.
Data are taken from Ref.~\cite{PRD19p2678} ($\odot$)
and Ref.~\cite{NPB296p493} ($\bullet$).
}
\label{KpiS}
\end{figure}

Recent studies in unquenched LQCD confirm the results of much older
coupled-channel analyses, such as a calculation \cite{PRL111p222001} of
the $D_{s0}^{*}(2317)$ meson as a scalar $c\bar{s}$ state with an important
$DK$ component \cite{PRL91p012003}. However, limitations due to problems in
dealing with very broad and overlapping resonances, as well as the physical
pion mass, have not yet allowed unquenched LQCD to identify the $K_0^*(800)$
resonance \cite{PRD86p054508}, predicted long ago in the coupled-channel
model of Ref.~\cite{ZPC30p615}.
In Fig.~\ref{KpiS} we show the results of HORSE
for $S$-wave $K\pi$ scattering
cross sections in the isospin $I=1/2$ channel \cite{ZPC30p615}.
For small values of the overall coupling constant $\lambda$,
one finds in the model narrow peaks at the harmonic-oscillator masses
of $u\bar{s}$ states.
However, for the model value $\lambda =1$ one obtains
a fair agreement between model \cite{ZPC30p615}
and experiment \cite{PRD19p2678,NPB296p493}.
Inspection of the model's transition amplitude reveals
an additional resonance pole at $s=\left( 0.772-0.281 i\right)^{2}$,
besides the poles that correspond to the harmonic-oscillator states.
The former resonance pole is absent for small $\lambda$,
hence dynamically generated by the interaction between
the quark-antiquark and two-meson configurations.
At $s=\left( 2.04-0.15 i\right)^{2}$ we find a further
dynamically generated resonance pole,
which is unexpected since both data and model
show a minimum in the $S$-wave $K\pi$ scattering
cross sections in the isospin $I=1/2$ channel
at the corresponding invariant mass.
The latter phenomenon underlines once more the need
for high-statistics data.

It is basically unknown
that a further consequence of ha\-dro\-nic com\-po\-si\-te\-ness
is the appearance of non-resonant threshold en\-han\-ce\-ments.
A theoretical model for threshold en\-han\-ce\-ments
in ha\-dro\-nic production amplitudes,
based on quark-antiquark pair creation,
was formulated in Ref.~\cite{AP323p1215}
and further developed in Refs.~\cite{EPL81p61002,EPL84p51002}.
This model shows that one must expect non-resonant en\-han\-ce\-ments
in the amplitudes just above pair-creation thresholds, which,
in the case of stable hadrons,
are accompanied by a clear minimum at threshold,
as observed in experiment for the process $e^{+}e^{-}\to b\bar{b}$,
measured and analysed by the BaBar Col\-la\-bo\-ra\-tion
\cite{PRL102p012001}.
As also remarked by BaBar in their paper,
the large statistics and the small energy steps of the
scan make it possible to clearly observe the dips
at the opening of the thresholds corresponding
to the $B\bar{B}^{\ast}+\bar{B}B^{\ast}$
and $B^{\ast}\bar{B}^{\ast}$ channels.
However, experimental evidence of this phenomenon is scarce,
since it needs event counts with high statistics and good resolution.
Nevertheless, in some cases signals, albeit often feeble,
can be seen in experimental data for ha\-dro\-nic production
\cite{PRD80p074001}.

In Sect.~\ref{Threnh} we study some examples of possibly observed
threshold enhancements, thereby stressing the need for
more accurate data.
The possibly observed superlight boson is discussed
in Sect.~\ref{E38boson}.

\section{Threshold enhancements}
\label{Threnh}

In Ref.~\cite{AP323p1215} the generic relation
\begin{equation}
P=\Imag{Z}+TZ
\label{production}
\end{equation}
between two-particle scattering ($T$) and production ($P$) amplitudes
is studied in a microscopic multichannel model for meson-meson scattering
with coupling to confined quark-antiquark channels.
The amplitude $T$ in expression (\ref{production}) is supposed
to contain the resonance poles which occur in scattering,
whereas $Z$ is a smooth function of invariant mass.
Threshold enhancements occur in production amplitudes
as a consequence of the shape of $\Imag{Z}$,
which in the ideal case of no further nearby thresholds
rises sharply just above threshold.
For larger invariant masses $\Imag{Z}$ first reaches a maximum
and then falls off exponentially.
As a consequence, production amplitudes show non-resonant yet
resonant-like enhancements just above threshold.

\subsection{Electron-positron annihilation into open-bottom pairs}
\label{epemBB}

\begin{figure}[htbp]
\begin{center}
\begin{tabular}{ccc}
\includegraphics[height=90pt]{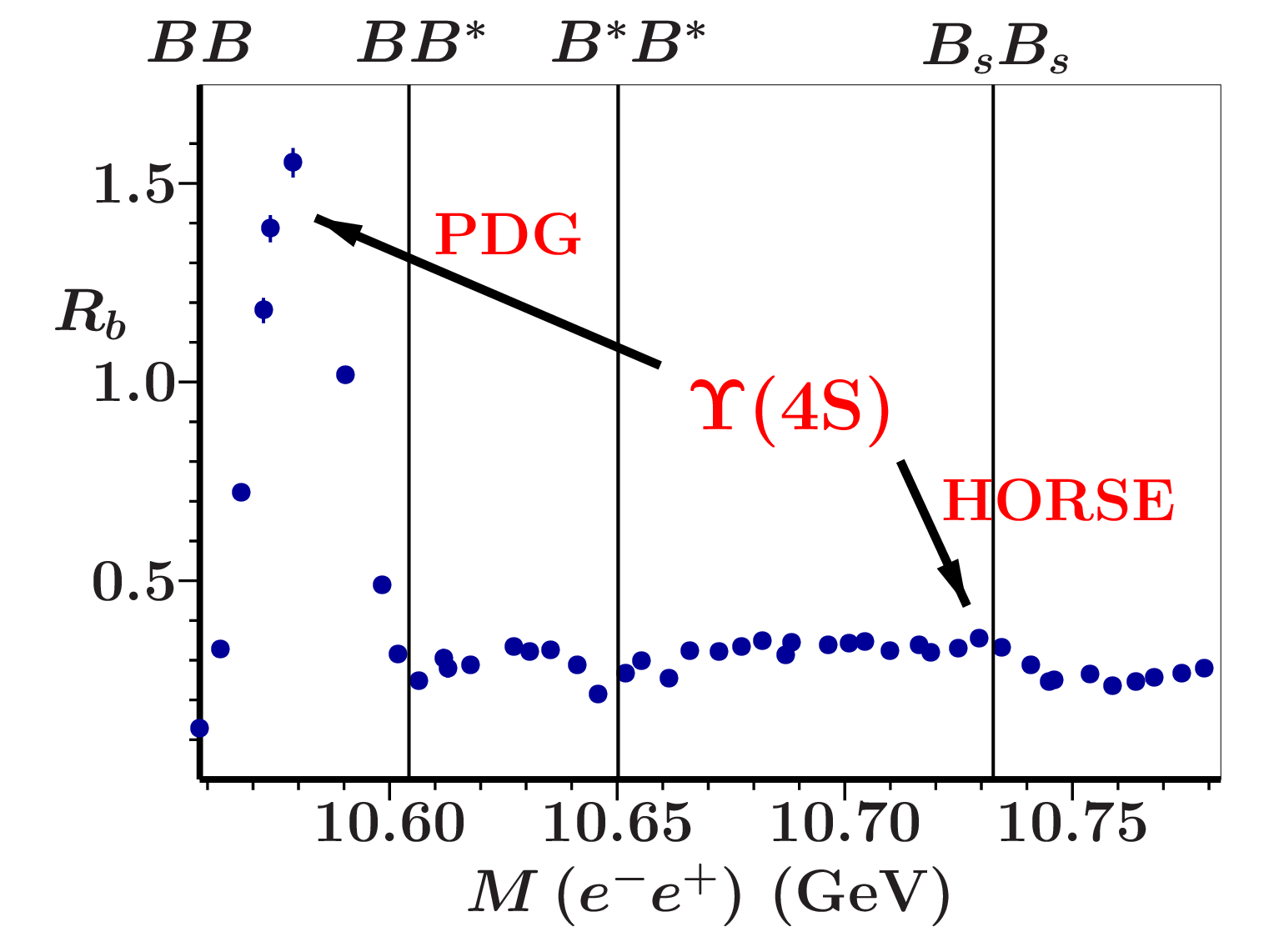} &
\includegraphics[height=90pt]{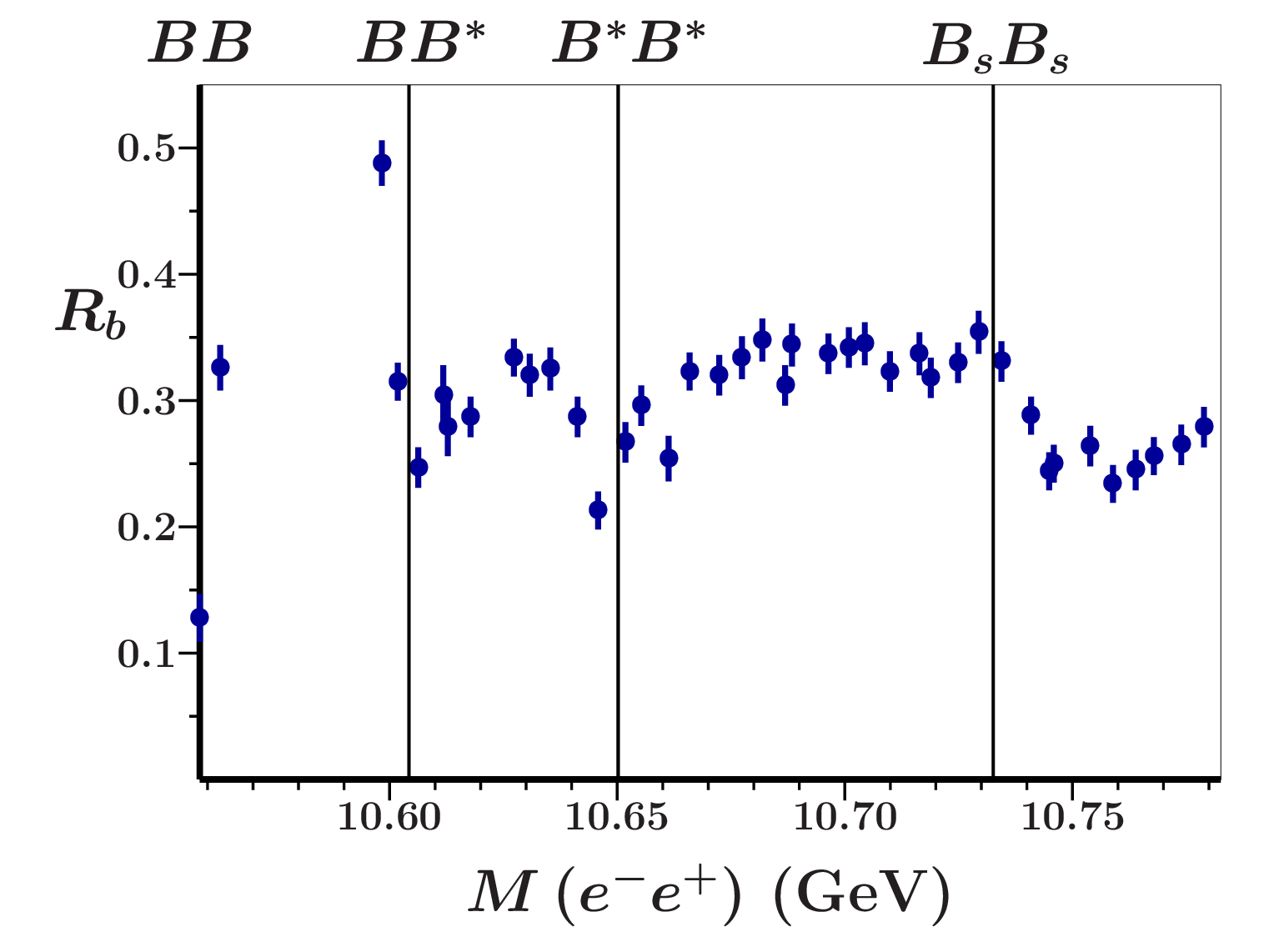} &
\includegraphics[height=90pt]{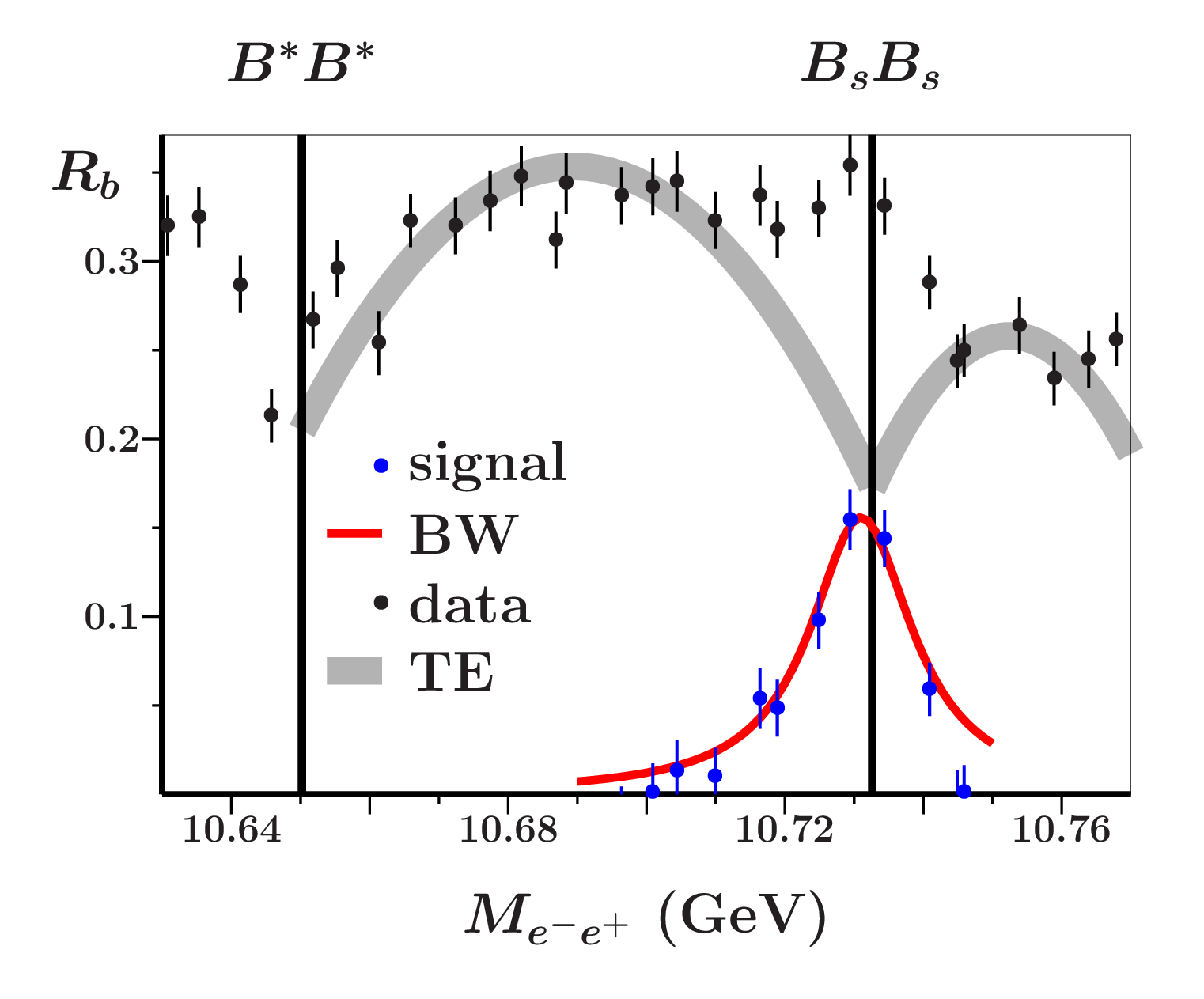}\\
(a) & (b) & (c)
\end{tabular}
\end{center}
\caption[]{$R_{b}$ as measured by BaBaR \cite{PRL102p012001}
in electron-positron annihilation;
(a): Full data, (b): without the enhancement at 10.58 GeV,
(c): near the $B_{s}B_{s}$ threshold.
In (c) a Breit-Wigner (BW) approximation is shown for the difference
signal between the data and threshold-enhancement (TE) structures.
}
\label{ups4S}
\end{figure}

In Fig.~\ref{ups4S} we show data on hadron production
in electron-positron annihilation, published by the BaBar Collaboration
in Ref.~\cite{PRL102p012001}.
In Fig.~\ref{ups4S}a we show the full data in the mass region
contained between the $BB$ and $BB^{\ast}$ threshold,
whereas in Fig.~\ref{ups4S}b we concentrate on the data
above the $BB^{\ast}$ threshold.
We observe in Fig.~\ref{ups4S} three threshold enhancements
at the opening of $BB$, $BB^{\ast}$ and $B^{\ast}B^{\ast}$,
which most likely are non-resonant.
The BaBar Collaboration remarks in Ref.~\cite{PRL102p012001}
that large statistics and the small energy steps of the scan
make it possible to clearly observe dips at the opening
of the thresholds corresponding to $BB^{\ast}$ and $B^{\ast}B^{\ast}$.
The shapes of the various threshold enhancements do not show
the just mentioned exponential tails,
as a result of interference between the various configurations
$BB$, $BB^{\ast}$, $B^{\ast}B^{\ast}$, and so on.
Such configurations are of course also present below threshold,
though not capable of materialising.

In Fig.~\ref{ups4S}c we show the difference signal
near the opening of $B_{s}B_{s}$,
which results from subtracting
the $B^{\ast}B^{\ast}$ and $B_{s}B_{s}$ threshold enhancements
from the full data.
A Breit-Wigner approximation to the difference signal
clearly shows the shape of a resonance.
According to us, this is the $\Upsilon (4S)$ state,
rather than the large enhancement at 10.58 GeV and also
in fair agreement with the predictions
in Refs.~\cite{PRD21p772,PRD27p1527}.

\begin{figure}[htbp]
\centering
\sidecaption
\includegraphics[width=180pt,clip]{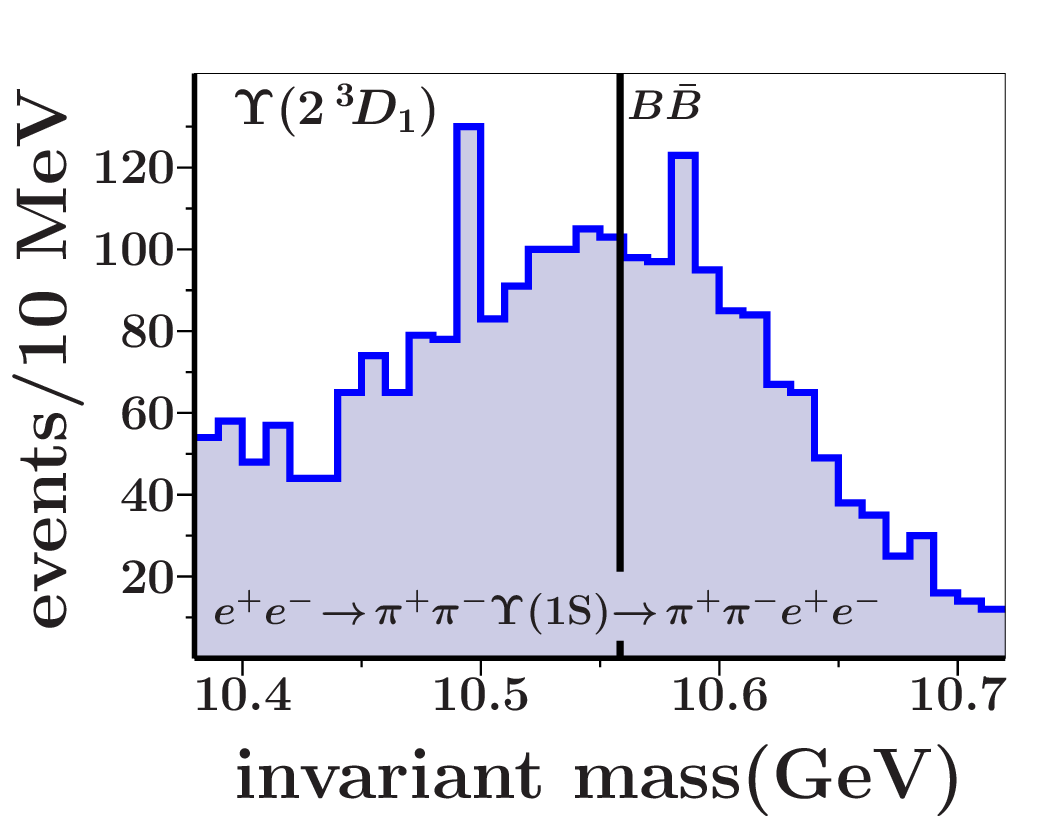}
\caption[]{
Narrow signal obtained for the
$\Upsilon\left( 2\,{}^{3\!}D_{1}\right)$ resonance
at 10495$\pm$5 MeV \cite{ARXIV10094097}.
Data are taken from Ref.~\cite{PRD78p112002}.
For completeness, we have also indicated
the $B\bar{B}$ and $B\bar{B}^{\ast}+\bar{B}B^{\ast}$ thresholds.
}
\label{2Dbb}
\end{figure}

The reason why the $BB$ threshold enhancement is much more
pronounced than those for the $BB^{\ast}$ and $B^{\ast}B^{\ast}$ channels
can be explained by the presence of a strong resonance pole
in the scattering amplitude on the real invariant-mass axis
just below the $BB$ threshold at about 10.5 GeV.
Data on $e^{+}e^{-}$ $\to$ $\pi^{+}\pi^{-}\Upsilon (2S,1S)$
$\to$ $\pi^{+}\pi^{-}e^{+}e^{-}$
published by the BaBar Collaboration \cite{PRD78p112002}
are further analysed in Ref.~\cite{ARXIV10094097}.
In Fig.~\ref{2Dbb} we depict the signal at 10495$\pm$5 MeV,
which was obtained in Ref.~\cite{ARXIV10094097} for the
$\Upsilon\left( 2\,{}^{3\!}D_{1}\right)$ resonance
and which is in good agreement with the prediction
in Ref.~\cite{PRD21p772}.
The signal, which is even more prominent than
the enhancement at 10.58 GeV, implies a strong resonance pole
on the real axis in the complex invariant-mass plane.
It is this pole that mainly contributes to the
threshold enhancement above the $B\bar{B}$ threshold.

\subsection{Electron-positron annihilation into open-charm pairs}
\label{epemDD}

Further threshold enhancements are discussed in
Ref.~\cite{PRD80p074001}.
Here we will concentrate on the reaction of $e^{+}e^{-}$ annihilation
into open-charm pairs, which can be observed
at and above the $D\bar{D}$ threshold.
We assume here that the reaction takes place via a photon
and the $c\bar{c}$ propagator,
through the creation of a light $q\bar{q}$ pair.
However, many competing configurations may be formed,
increasing in number for higher invariant masses.
Furthermore, it seems we may conclude from experiment
that stable open-charm hadrons have more probability to be
formed near threshold, where kinetic energy is almost zero.
Hence, if it were not for phase space and the centrifugal barrier,
$D\bar{D}$ pairs would be produced most likely just above threshold.

For total invariant mass below but close to the $DD^{\ast}$ threshold,
we assume that the probability of $D\bar{D}$ creation
is already reduced because of the non-vanishing probability of creating a
virtual $DD^{\ast}$ pair.
Just above the $DD^{\ast}$ threshold,
$D\bar{D}$ creation decreases rapidly.
The expected corresponding non-resonant contribution
of the production amplitude should thus exhibit this feature.
In this respect, an important observation was published
by the BES Collaboration in Ref.~\cite{ARXIV08070494}.
To our knowledge, the BES Collaboration was the first to discover
that the $\psi (3770)$ cross section is built up
by two different amplitudes, viz.\ a relatively broad signal and a
rather narrow $c\bar{c}$ resonance.
For the narrow resonance, which probably corresponds to
the well-established $\psi (1D)(3770)$,
the BES Collaboration measured a central resonance position
of $3781.0\pm 1.3\pm 0.5$ MeV
and a width of $19.3\pm 3.1\pm 0.1$ MeV (their solution 2).
If the latter parameters are indeed confirmed,
it would be yet another observation
of a quark-antiquark resonance width
that is very different from the world average
($87.04\pm 0.33$ MeV \cite{CNPC38p090001} in this case),
after a similar result was obtained by the BaBar Collaboration
in Ref.~\cite{PRL102p012001}, for $b\bar{b}$ resonances.
Concerning the broader structure, the BES Collaboration indicated, for
their solution 2, a central resonance position of $3762.6\pm 11.8\pm 0.5$ MeV
and a width of $49.9\pm 32.1\pm 0.1$ MeV.
The signal significance for the new enhancement is $7.6\sigma$
(solution 2). It was explained as a possible diresonance
\cite{PRD78p116014} or heavy molecular state \cite{ARXIV08080073}.
Here, we assume that the broader structure is most likely
caused by the non-resonant contribution
to the production amplitude given in Eq.~(\ref{production}),
thus lending further support to the idea that
the $\psi (1D)(3770)$ enhancement consists of two distinct signals.

\begin{figure}[htbp]
\centering
\sidecaption
\includegraphics[width=180pt,clip]{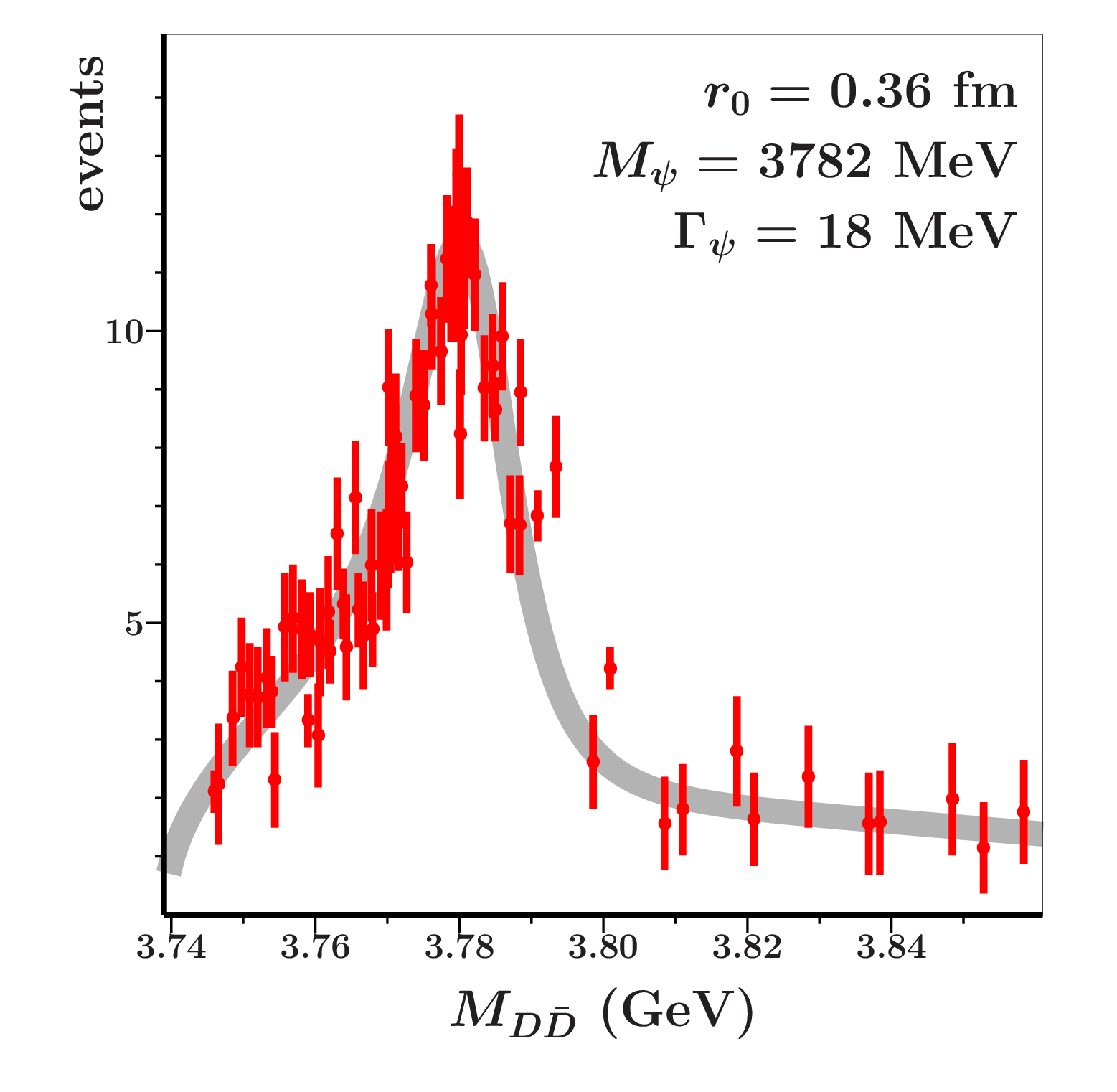}
\caption[]{\small
Comparison of experimental data obtained by the BES Collaboration
\cite{ARXIV08070494}
and the cross section resulting from Eq.~(\ref{production}),
by the use of expressions (\ref{Zfun}) and (\ref{Tfun}).
The non-resonant contribution
dominates for larger relative $D\bar{D}$ momentum $p$.
The BW parameters of the $c\bar{c}$ resonance are
$M\left(\psi (1D)\right) =3.782$ GeV and
$\Gamma\left(\psi (1D)\right) =18$ MeV.
}
\label{bes}
\end{figure}

The data shown in Fig.~\ref{bes} are fitted by assuming
\begin{equation}
Z={\cal N}pr_{0}\exp\left\{ -\left( pr_{0}\right)^{2}\right\}
e^{\textstyle i\phi}
\;\;\;\;\;\;\;\;\;\;\;\;
\left( 4p^{2}=s-4M^2_{D}\right)
\label{Zfun}
\end{equation}
for the $Z$ function in Eq.~(\ref{production}),
and
\begin{equation}
T=\fnd{\Gamma_{\psi}/2}{\sqrt{s}-M_{\psi}+i\Gamma_{\psi}/2}
\label{Tfun}
\end{equation}
for the scattering amplitude in Eq.~(\ref{production}).
The theoretical curve in Fig.~\ref{bes} corresponds to
$M_{\psi}=3782$ MeV, $r_{0}=0.36$ fm, $\Gamma_{\psi}=18$ MeV
and $\phi =0.7\pi$.
The overall normalisation factor ${\cal N}$ in Eq.~(\ref{Zfun})
is adjusted to the data.

Note the sharp interference dip at about 3.81 GeV,
which can also be observed in data published by the BaBar Collaboration
\cite{PRD79p092001}, but with even less statistics.

\subsection{\bm{D^{0}K^{-}} invariant-mass distribution}
\label{D0Km}

When the $D_{sJ}^{\ast}(2860)^{+}$ meson was discovered
by the BaBar Collaboration \cite{PRL97p222001},
only the $DK$ decay mode was detected, which made an
assignment as a radially excited scalar $c\bar{s}$ meson plausible
\cite{PRL97p202001,PLB647p159,EPJC50p617},
though other configurations such as a $3^{-}$ state
\cite{PLB642p48,EPJC50p617,NPPS186p363} could not be excluded.
For a discussion of additional options, see Ref.~\cite{NPA856p88}.
The later observed $D^{\ast}K$ mode \cite{PRD80p092003}
at first seemed to exclude the $0^{+}$ scenario for
the $D_{sJ}^{\ast}(2860)^{+}$ resonance.
But the true situation may be subtler, involving two overlapping
resonances, viz.\ one scalar $2\,{}^{3\!}P_{0}$ and
one tensor $2\,{}^{3\!}P_{2}$ charm-strange meson \cite{PRD81p118101}.
However, in a recent study the LHCb Collaboration finds
overlapping spin-1 and spin-3 resonances at a mass 2.86 GeV,
by analysing the resonant $D^{0}K^{-}$ substructure
of $B_{s}^{0}\to\bar{D}^{0}K^{-}\pi^{+}$ decays
with the technique of Dalitz-plot analysis \cite{PRD90p072003,ARXIV14077574}.
Here we will study the latter result.

The angular analysis of the $D_{sJ}^{\ast}(2860)^{-}\to D^{0}K^{-}$ decays
supports natural parity for the resonance, i.e.,
$J^{P}=0^{+}$, $1^{-}$, $2^{+}$, $3^{-}$, $\dots$.
Its mass spectrum for harmonic-oscillator confinement \cite{PRD27p1527}
is given in Table~\ref{natpar}.
\begin{table}[htbp]
\begin{center}
\begin{tabular}{||c||c||}
\hline\hline & \\ [-7pt]
Mass (GeV) & States\\
& \\ [-7pt]
\hline & \\ [-7pt]
3.305 & $3^{3}\! P_{0}$ $3^{3}\! P_{2}$ $2^{3}\! F_{2}$
$2^{3}\! F_{4}$ $1^{3}\! H_{4}$ $1^{3}\! H_{6}$\\
3.115 & \bm{3^{3}\! S_{1}} $2^{3}\! D_{1}$ $2^{3}\! D_{3}$
$1^{3}\! G_{3}$ $1^{3}\! G_{5}$\\
2.925 & \bm{2^{3}\! P_{0}} \bm{2^{3}\! P_{2}}
\bm{1^{3}\! F_{2}} \bm{1^{3}\! F_{4}}\\
2.735 & $2^{3}\! S_{1}$ $1^{3}\! D_{1}(2709)$
$1^{3}\! D_{3}$\\
2.545 & $1^{3}\! P_{0}$ $1^{3}\! P_{2}(2572)$\\
2.355 & $1^{3}\! S_{1}(2110)$\\
\hline\hline
\end{tabular}
\end{center}
\caption[]{\small $D_{s}$ natural-parity HO states and their masses.
When experimental central resonance masses are measured,
those are shown in brackets.
The states in boldface are discussed in the text.}
\label{natpar}
\end{table}
We observe at a glance that, besides the well-established
$1^{3}\! S_{1}(2110)$ and $1^{3}\! P_{2}(2572)$ resonances,
most of the predicted states have not even been observed
or their quantum numbers are still to be determined.
In Ref.~\cite{ARXIV10112360} the assignment $2^{3}\! S_{1}$
for the resonance at 2.71 GeV is disputed.
As can be read from Table~\ref{natpar},
we expect four resonances in the mass interval 2.735--2.925 GeV,
namely $2^{3}\! P_{0}$, $2^{3}\! P_{2}$, $1^{3}\! F_{2}$ and $1^{3}\! F_{4}$,
while out of the five predicted resonances
in the mass interval 2.925--3.115 GeV
we expect for $3^{3}\! S_{1}$ the larger mass shift
and thus to come out closer to 2.925 GeV than the other four.

\begin{figure}[htbp]
\begin{center}
\begin{tabular}{ccc}
\includegraphics[height=100pt]{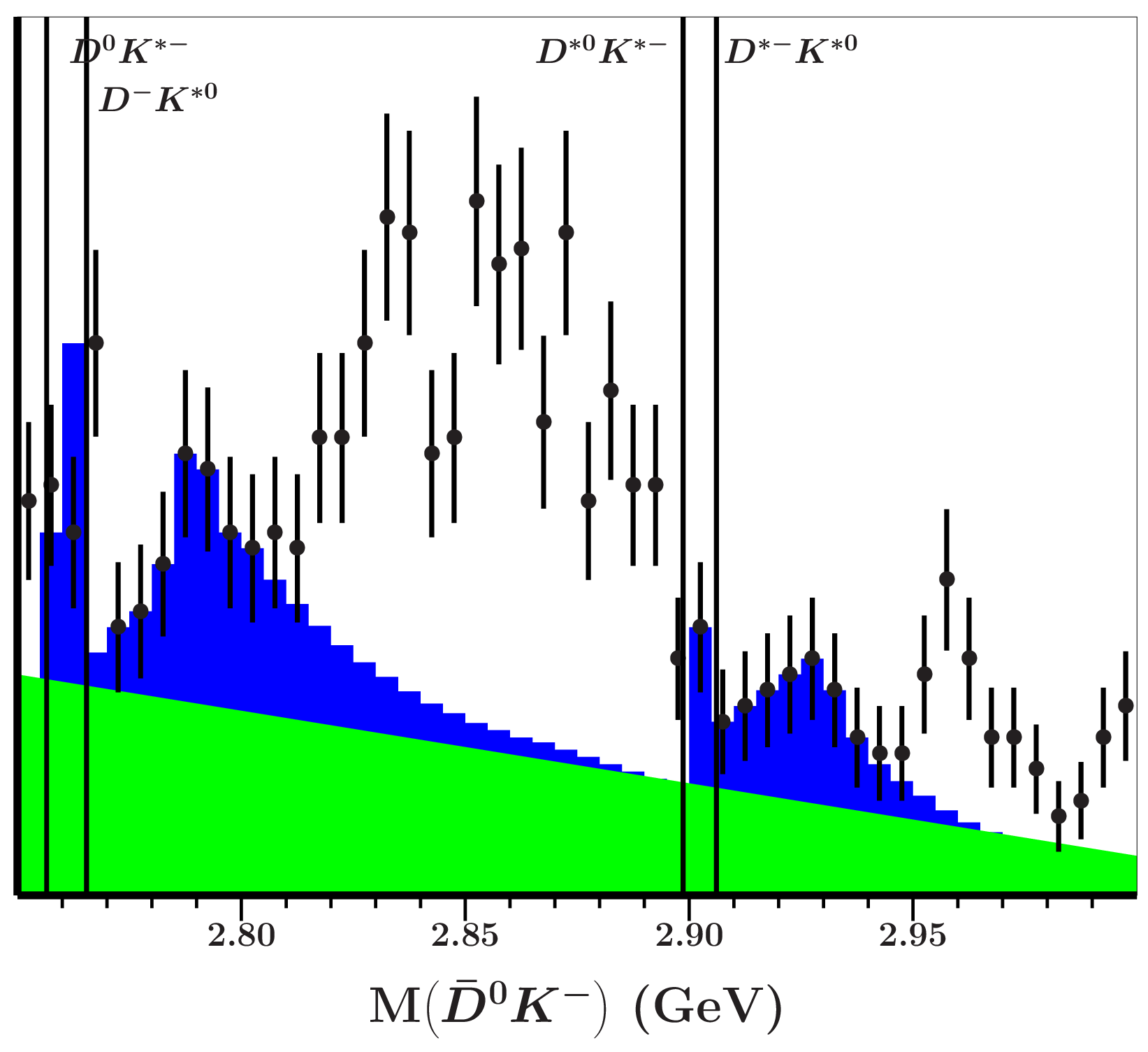} &
\includegraphics[height=100pt]{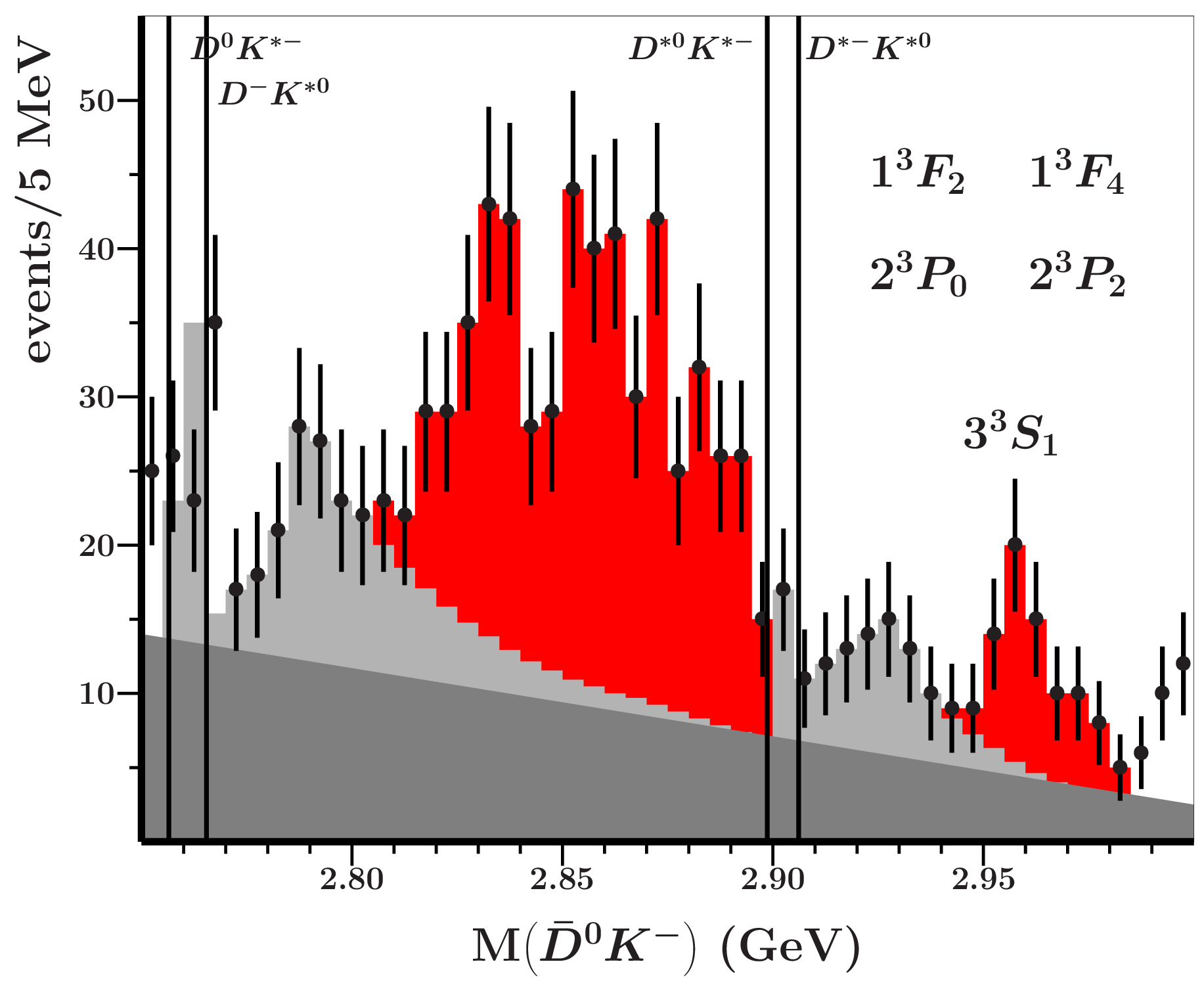} &
\includegraphics[height=100pt]{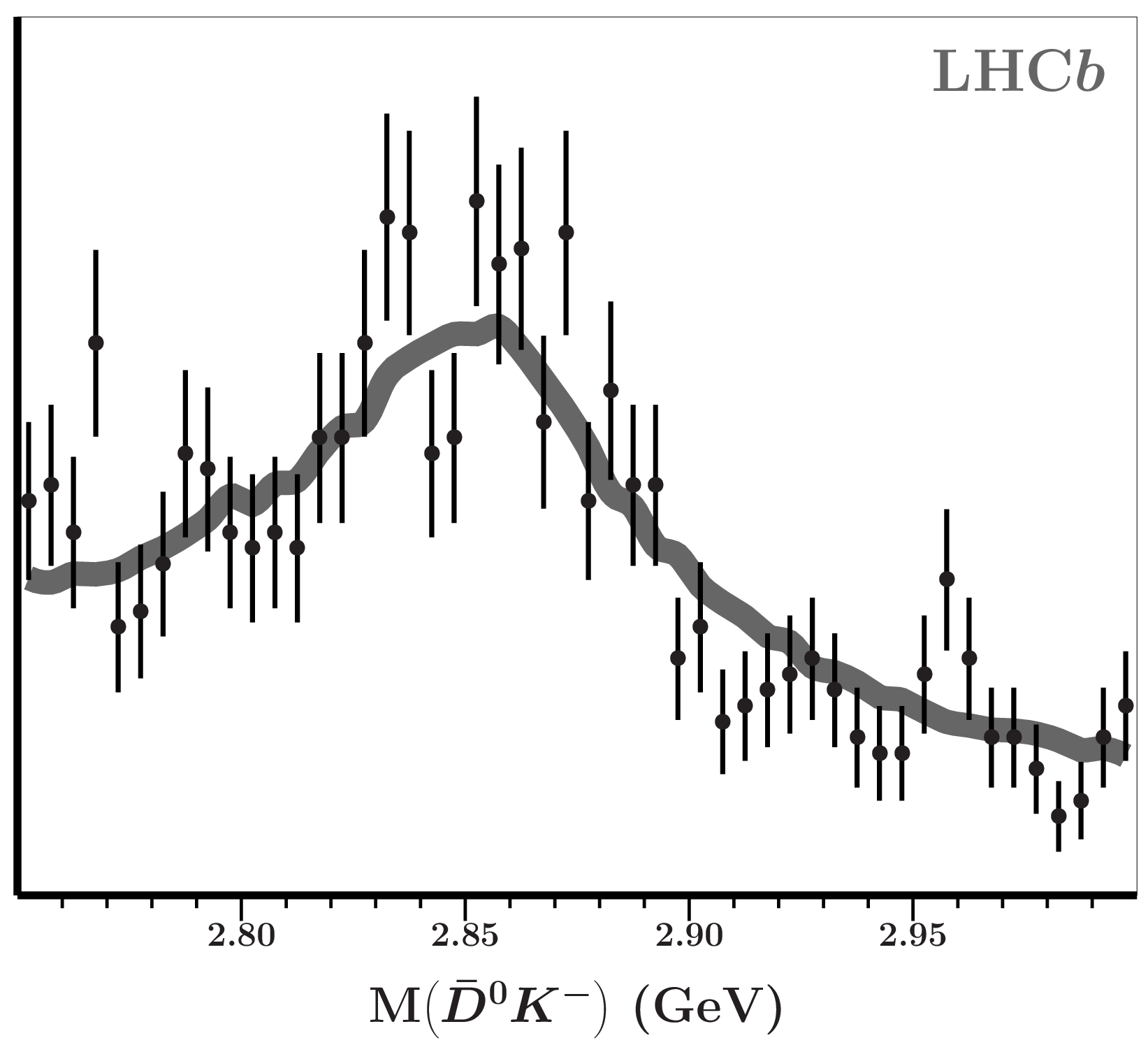}\\
(a) & (b) & (c)
\end{tabular}
\end{center}
\caption[]{$D^{0}K^{-}$ mass distribution in
$B_{s}^{0}\to\bar{D}^{0}K^{-}\pi^{+}$ decay
\cite{PRD90p072003,ARXIV14077574};
(a): Our suggestion for background and threshold enhancements,
(b): Highlighting the remaining structures,
(c): LHCb representation of the data.
}
\label{LHCbD0K}
\end{figure}

In Fig.~\ref{LHCbD0K} we show the $D^{0}K^{-}$ mass distribution
in the mass interval 2.75--3.0 GeV as published by LHCb
\cite{PRD90p072003,ARXIV14077574}.
Besides an estimate for a possible background,
the threshold enhancements at
the openings of the $D^{0}K^{\ast -}$/$D^{-}K^{\ast 0}$ channels
and the $D^{\ast 0}K^{\ast -}$/$D^{\ast -}K^{\ast 0}$ channels
are shown in Fig.~\ref{LHCbD0K}a.
The remaining signal is highlighted in Fig.~\ref{LHCbD0K}b.
The $3^{3}\! S_{1}$ assignment for the peak at 2.96 GeV
seems to be in good agreement with our expectation from the HO
value of 3115 MeV and a mass shift of roughly 160 MeV due to meson loops.
The signal peaking at about 2.86 GeV could consist
of 2, 3 or 4 overlapping resonances.
This issue can only be solved with much better statistics
and bin sizes that do not exceed 1.0 MeV.
The LHCb solution for the data shown here
is depicted in Fig.~\ref{LHCbD0K}c.

\subsection{Weak substructure?}
\label{weak}

This section is devoted to the recently observed
(pseudo)scalar, most probably scalar
\cite{PRL110p081803,PLB726p120},
enhancement in the 120--135 GeV interval
\cite{PLB716p1,PLB716p30}
in the light of non-resonant threshold en\-han\-ce\-ments.

\begin{figure}[htbp]
\begin{center}
\begin{tabular}{ccc}
\includegraphics[height=100pt]{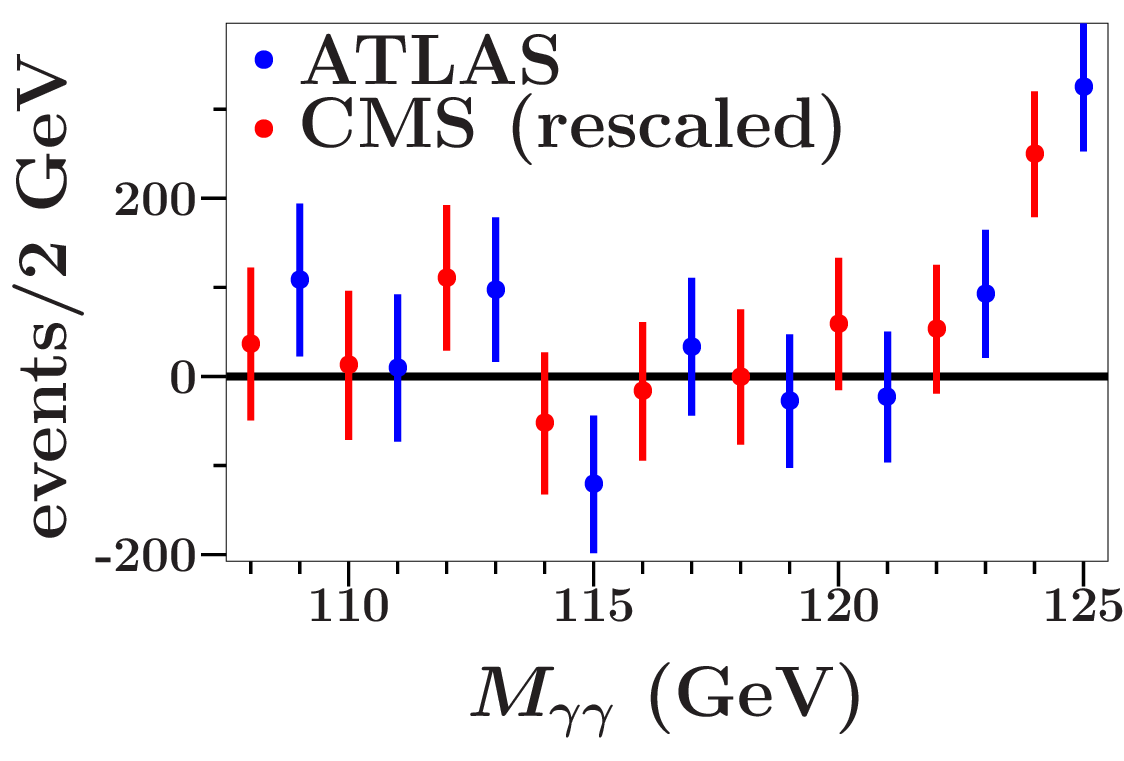} &
\includegraphics[height=100pt]{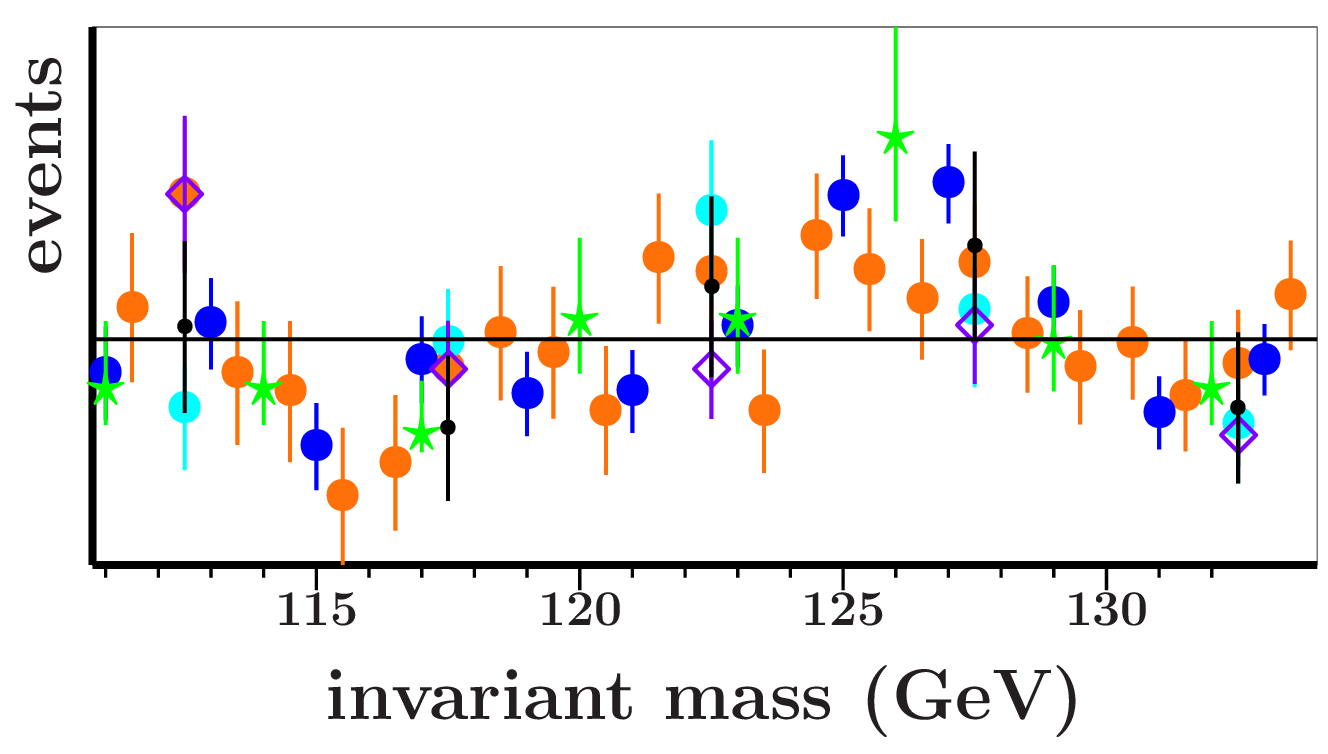}\\
(a) & (b)
\end{tabular}
\end{center}
\caption[]{(a): Diphoton signals after background subtraction,
published by the CMS Collaboration \cite{EPJC74p3076}
({\color{red}$\bullet$}) and the ATLAS Collaboration
\cite{ARXIV13053315} ({\color{blue}$\bullet$});
(b): Diphoton signals published
by the CMS Collaboration \cite{CMSPASHIG-13-016}
({\definecolor{tmpclr}{rgb}{1.000,0.440,0.030}{\color{tmpclr}$\bullet$}})
and the ATLAS Collaboration
\cite{ARXIV13053315} ({\color{blue}$\bullet$}),
four-lepton signals published
by the CMS Collaboration \cite{ARXIV13125353}
({\color{green}$\star$})
and the ATLAS Collaboration \cite{ARXIV13053315}
({\definecolor{tmpclr}{rgb}{0.000,1.000,1.000}{\color{tmpclr}$\bullet$}}),
invariant-mass distributions for
$\tau\tau$ in $e^{+}e^{-}\to\tau\tau (\gamma )$
({\definecolor{tmpclr}{rgb}{0.500,0.000,1.000}{\color{tmpclr}$\diamond$}})
and
$\mu\mu$ in $e^{+}e^{-}\to\mu\mu (\gamma )$ ({$\bullet$})
published
by the L3 Collaboration \cite{PLB479p101}.
}
\label{dip115}
\end{figure}

In Ref.~\cite{EPJC74p3076} the CMS Collaboration
collected diphoton events in proton-proton collisions
corresponding to integrated luminosities of 5.1 fb$^{-1}$
at a centre-of-mass energy of 7 TeV, and 19.7 fb$^{-1}$ at 8 TeV.
The ATLAS Collaboration published similar results \cite{ARXIV13053315}
based on data samples corresponding to integrated luminosities
of up to 20.7 fb$^{-1}$ at 8 TeV, and 4.6 fb$^{-1}$ at 7 TeV.
Those data are collected in Fig.~\ref{dip115}a.

Besides the bump at about 125 GeV, one observes a clear dip
in the diphoton invariant-mass distribution at about 115 GeV.
Its signal is rather similar to the threshold dips observed
by the BaBar Collaboration \cite{PRL102p012001}
and which are depicted in Fig.~\ref{ups4S}.
Hence, could there exist a threshold at 115 GeV
for particle-antiparticle production?
If so, the masses of the particles should be about 57.5 GeV each.

Composite heavy gauge bosons and their spin-zero partners,
the latter with a mass in the range 50--60 GeV,
were considered long ago \cite{PLB135p313}
and studied in numerous works (see e.g.\ Refs.\
\cite{PLB141p455,PRAMANA23p607,PRD36p969,NCA90p49,PRD39p3458,PRL57p3245}).
To date, no experimental evidence of their existence has been reported.
More recently the interest in Weak substructure has revived
\cite{ARXIV12074387,ARXIV12105462,ARXIV13076400,ARXIV13040255,PRD90p035012}.

In Fig.~\ref{dip115}b we collect invariant-mass distributions for
a variety of two-particle production processes, namely
diphoton signals after background subtraction
published by the CMS \cite{CMSPASHIG-13-016}
and ATLAS Collaborations \cite{ARXIV13053315},
four-lepton signals published
by the CMS \cite{ARXIV13125353}
and ATLAS Collaborations \cite{ARXIV13053315},
and for $\tau\tau$ in $e^{+}e^{-}\to\tau\tau (\gamma )$
and $\mu\mu$ in $e^{+}e^{-}\to\mu\mu (\gamma )$
published by the L3 Collaboration in Ref.~\cite{PLB479p101}.
The data shown in Fig.~\ref{dip115}
indeed seem to suggest that the total signal
in the 115--133 GeV energy interval is built up
by two different amplitudes, viz.\ a broad non-resonant signal
and a resonance at about 125 GeV.

In order to proceed let us assume that a (pseudo-)scalar partner
of the $Z$-boson exists, say a $\tilde{Z}(57)$.
By its quantum numbers we may expect that the $\tilde{Z}(57)$
does not show up in dilepton cross sections,
or at least not as a very clear resonance.
But it could be visible in diphoton data.
Indeed, a promising analysis of the L3 Collaboration
\cite{PLB345p609} does show that $Z(91)\to\gamma\tilde{Z}(57)$
transitions cannot be excluded a priori.
In their analysis, the L3 Collaboration searched for anomalous
$Z\to\gamma\gamma\gamma$ events with the L3 detector at LEP
and concluded that no significant deviations were observed
from the $e^{+}e^{-}\to\gamma\gamma\gamma$ events expected by QED.
We do not contest here the conclusion of the L3 Collaboration,
since with a sample of 87 candidate events,
against $(76.3\pm 2.8)$ events expected from QED processes,
one can hardly expect
sufficient resolution and statistics for evidence of the existence
of a hard-to-be-detected pseudoscalar partner of the $Z$ boson.
Nevertheless, it is justified to pay here closer attention to the result
obtained by the L3 Collaboration.

\begin{figure}[htbp]
\begin{center}
\begin{tabular}{cc}
\includegraphics[height=150pt]{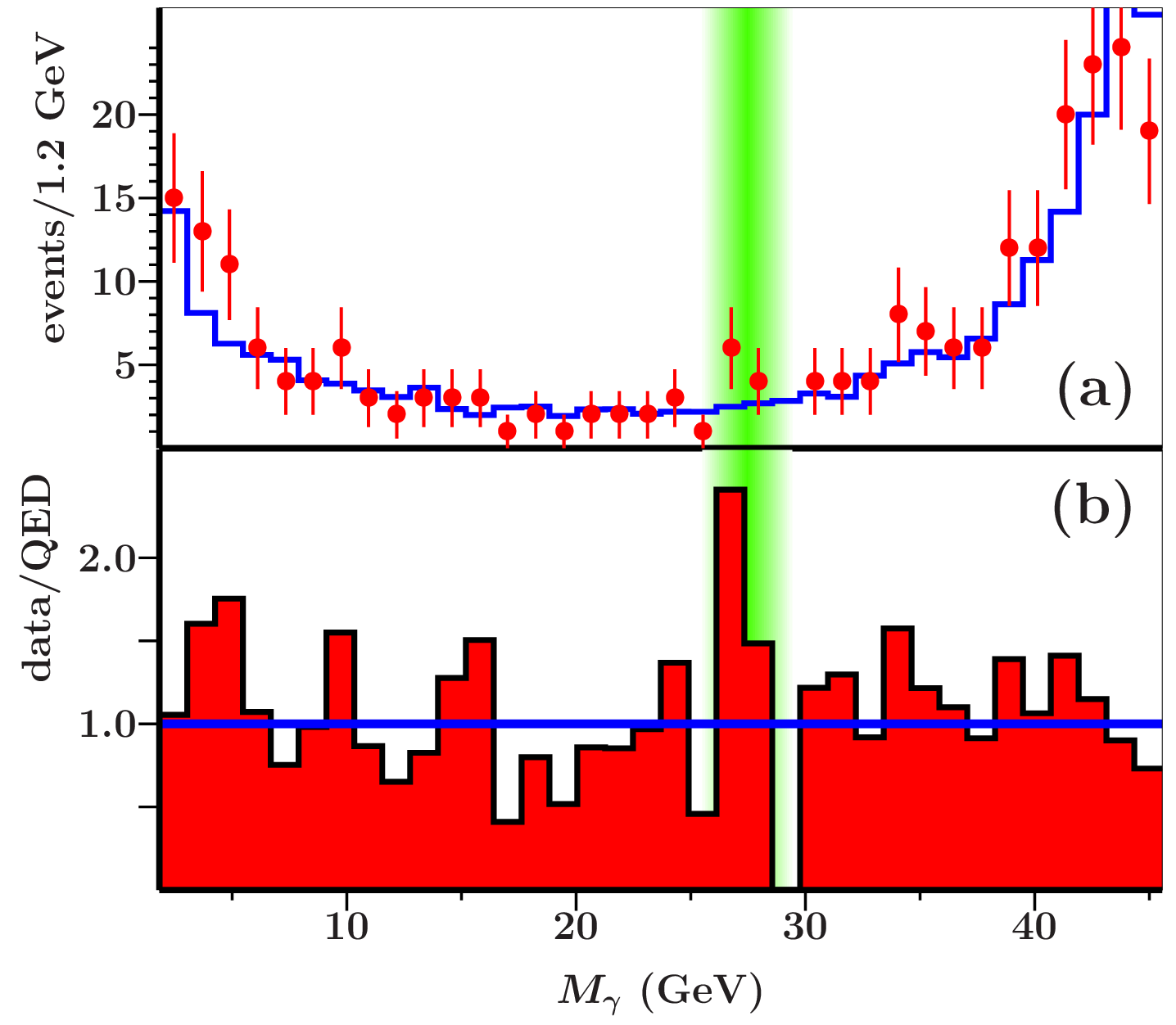} &
\includegraphics[height=150pt]{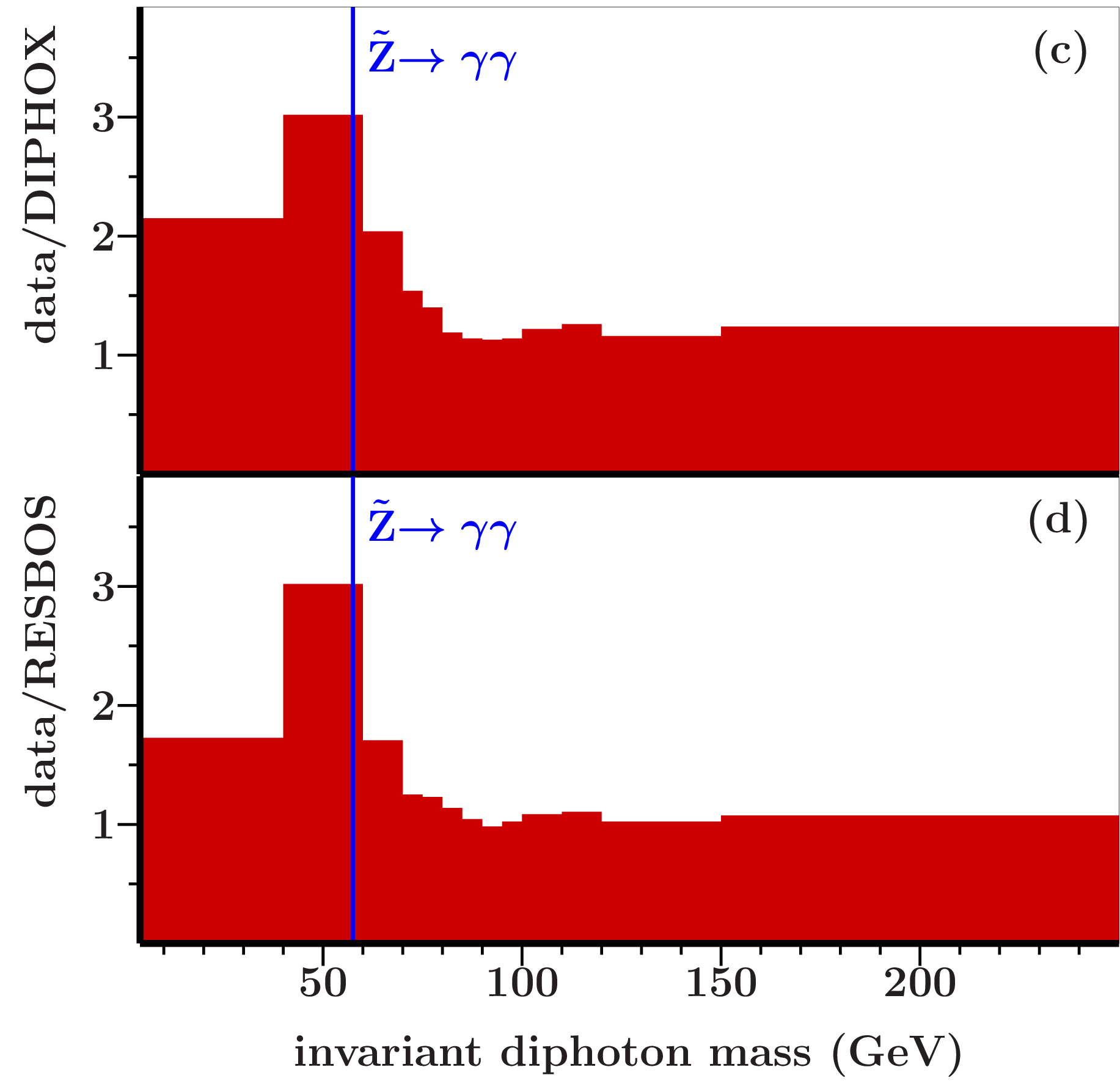}
\end{tabular}
\end{center}
\caption[]{
(a): Experimental data for the three one-photon CM energies
of the candidate $Z\to\gamma\gamma\gamma$ events
obtained by the L3 Collaboration \cite{PLB345p609},
assuming $\sqrt{s}=M_{Z}$.
The histogram was obtained by the L3 Collaboration
from a Monte-Carlo simulation for the expected number of events
predicted by QED.
With the green band we indicate where we expect
photons from the radiative process $Z(91)\to\gamma\tilde{Z}(57)$
for the case that $\tilde{Z}(57)$ has a mass of 57.5 GeV.
(b): The same data as shown in (a), but now measured events
divided by QED expected events.
(c, d): Measured over expected events for diphoton invariant-mass
distributions published by the CMS Collaboration \cite{CMSPASHIG-13-001},
for (c) DIPHOX and (d) RESBOS.
}
\label{1photonL3}
\end{figure}

In Fig.\ref{1photonL3}a/b we show the L3 data for
the three one-photon CM energies for each of the candidate events.
The L3 Collaboration collected the one-photon CM energies
as a function of $M_{\gamma}/\sqrt{s}$.
Here, we have converted that information into $M_{\gamma}$,
thereby assuming $\sqrt{s}=M_{Z}$.
With a green band we indicate where we expect
the photons from the radiative process $Z(91)\to\gamma\tilde{Z}(57)$.
We observe that most of the data agree well with the expectation from QED.
Nevertheless, be it a coincidence or not, in the mass region
where we expect a signal from $Z(91)\to\gamma\tilde{Z}(57)$ events,
we observe a small enhancement.
The latter can be better demonstrated by showing
the ratio of measured signal over QED prediction,
as also depicted in Fig.~\ref{1photonL3}b.
Here one clearly observes a modest enhancement
for exactly the expected $\tilde{Z}(57)$ mass of 57.5 GeV.

In diphoton invariant-mass distributions
published by the CMS Collaboration \cite{CMSPASHIG-13-001},
shown in Fig.~\ref{1photonL3}c for predictions
of a DIPHOX data simulator and
in Fig.~\ref{1photonL3}d for predictions of RESBOS,
one also observes an excess of three times more observed events
than predicted in the 40--60 GeV mass interval.
This could be in agreement with diphotons stemming from the decay
process $\tilde{Z}(57)\to\gamma\gamma$.

We have thus shown here that the described threshold
phenomena, observed for mesons, may also have an application to heavy
gauge bosons at high energies, where the weak interactions are in fact not
weak anymore.
Confirmation of such effects would indicate compositeness in this
sector of the Standard Model, too.
Moreover, it could offer an explanation for the
enhancement around 125 GeV seen by ATLAS and CMS at LHC, as a possible
alternative to a generally accepted Higgs-like particle of such a mass.
A consequence of this scenario would be the existence of gauge-boson partners,
of lower mass and with different quantum numbers, being either scalars or
pseudoscalars.
The LHC and LEP data we have presented above indeed hint at the
existence of such partners, namely a $\tilde{Z}(57)$ at 57.5 GeV.
Moreover, there are also indications \cite{ARXIV13047711}, albeit
feeble, of $Z$-like recurrences, viz.\ at about 210 and 240 GeV.
However, only much improved statistics and resolution
can settle the issue of a possible weak substructure
and its manifestation through new heavy bosons.

\section{Existence of a superlight scalar boson}
\label{E38boson}

In Ref.~\cite{ARXIV12021739}
a variety of indications were presented of the possible existence
of a light boson with a mass of about 38 MeV,
henceforth referred to as $E(38)$.
These indications amounted to a series of
low-statistics observations all pointing in the same direction,
and one high-statistics observation,
which might be interpreted as the discovery of the $E(38)$.

\subsection{Motivation}

In Ref.~\cite{NCA80p401} an $SO(4,2)$ conformally symmetric model
was proposed for strong interactions at low energies,
based on the observation \cite{THEFNYM7911,PRD30p1103,LNP211p331}
that confinement can be described
by an anti-De Sitter (aDS) background geometry.
The possibility of such a strategy had already been studied,
almost a century ago, by H.~Weyl \cite{AdP364p101},
who found that the dynamical equations of gauge theories
retain their flat-space-time form
when subject to a conformally-flat metrical field
instead of the usual Minkowski background.
The unification of electromagnetism and strong interactions
can be justified by the very subtle balance between these forces
in the nucleus, where just one neutron more or less
can make the difference between stability or instability.

Confinement of quarks and gluons was in Ref.~\cite{NCA80p401} modelled
by the introduction of two scalar fields, which
spontaneously break the $SO(4,2)$ symmetry down to
$SO(3,2)$ and $SO(3)\otimes SO(2)$ symmetry, respectively.
Moreover, a symmetric second-order tensor field was defined
that serves as the metric for flat space-time,
coupling to electromagnetism.
Quarks and gluons, which to lowest order
do not couple to this tensor field,
are confined to an aDS universe \cite{ARXIV07061887},
having a finite radius in the flat space-time.
This way, the model describes quarks and gluons, which
oscillate with a universal frequency ---
independent of the flavour mass --- inside a closed universe,
as well as photons, which freely travel through flat space-time.

The fields in the model of Ref.~\cite{NCA80p401}
comprise one real scalar field $\sigma$
and one complex scalar field $\lambda$.
Their dynamical equations were solved in Ref.~\cite{NCA80p401}
for the case that the respective vacuum expectation values,
given by $\sigma_{0}$ and $\lambda_{0}$,
satisfy the relation
\begin{equation}
\abs{\sigma_{0}}\gg \abs{\lambda_{0}}
\;\;\; .
\label{slvacua}
\end{equation}
A solution for $\sigma_{0}$ of particular interest
leads to aDS confinement, via the associated
conformally flat metric given by $\sigma\eta_{\mu\nu}$.
Furthermore, the only quadratic term
in the Lagrangian of Ref.~\cite{NCA80p401}
is proportional to $-\sigma^{2}\lambda^{\ast}\lambda$.
Hence, under the condition of relation (\ref{slvacua}),
one obtains, after choosing vacuum expectation values,
a light $\sigma$ field associated with confinement,
and a very heavy complex $\lambda$ field
associated with electromagnetism.
Here, we will study the --- supposedly light --- mass
of the scalar field that gives rise to confinement.

The conformally symmetric model of Ref.~\cite{NCA80p401}
in itself does not easily allow for interactions between hadrons,
as each hadron is described by a closed universe.
Therefore, in order to compare the properties of this model
to the actually measured cross sections and branching ratios,
the model has been further simplified,
such that only its main property survives,
namely its flavour-independent oscillations.
This way the full aDS spectrum is, via light-quark-pair creation,
coupled to the channels of two --- or more --- hadronic decay products
for which scattering amplitudes can be measured,
thus relating observed resonances to the aDS spectrum.

The aDS spectrum reveals itself through the structures
observed in hadronic invariant-mass distributions.
However, as we have shown in the past
(see Ref.~\cite{ARXIV10112360} and references therein),
there exists no simple relation between
enhancements in the experimental cross sections
and the aDS spectrum.
Nevertheless, this was studied in parallel, for mesons,
in a coupled-channel model
in which quarks are confined by
a flavour-independent harmonic oscillator
\cite{PRD21p772,PRD27p1527}.
Empirically, based on numerous data on mesonic resonances
measured by a large variety of experimental collaborations,
it was found \cite{AIPCP1374p421}
that an aDS oscillation frequency of $\omega =190$ MeV
agrees well with the observed results for
meson-meson scattering and meson-pair production
in the light \cite{ZPC30p615},
heavy-light \cite{PRL91p012003},
and heavy \cite{CNPC35p319,ARXIV10094097} flavour sectors,
thus reinforcing the strategy proposed in Ref.~\cite{NCA80p401}.

A further ingredient of the model for the description
of non-exotic quarkonia, namely the coupling
of quark-antiquark components
to real and virtual two-meson decay channels \cite{AP324p1620}
via $^{3\!}P_{0}$ quark-pair creation,
gives us a clue about the size of the mass of the $\sigma$ field.
For such a coupling it was found that the average radius $r_{0}$
for light-quark-pair creation in quarkonia can be described
by a flavour-independent mass scale, given by
$M=\frac{1}{2}\omega^{2}\mu r_{0}^{2}$,
where $\mu$ is the effective reduced quarkonium mass.
In earlier work, the value $\rho_{0}=\sqrt{\mu\omega}r_{0}=0.56$
\cite{PRD21p772,PRD27p1527} was used,
which results in $M=30$ MeV for the corresponding mass scale.
However, the quarkonium spectrum is not very sensitive
to the precise value of the radius $r_{0}$,
in contrast with the resonance widths.
In more recent work \cite{HEPPH0201006,PLB641p265},
slightly larger transition radii have been applied,
corresponding to values around 40 MeV for $M$.
Nevertheless, values of 30--40 MeV
for the flavour-independent mass $M$
do not seem to bear any relation
to an observed quantity for strong interactions.
However, in Refs.~\cite{ARXIV11021863,ARXIV12021739}
we have presented experimental evidence
for the possible existence of a quantum with a mass of about 38 MeV,
which in light of its relation to the $^{3\!}P_{0}$ mechanism
we suppose to mediate quark-pair creation.
Moreover, its scalar properties make it a perfect candidate
for the quantum associated with
the above-discussed scalar field for confinement.

\subsection{Interference effects}

\begin{figure}[ht]
\begin{center}
\begin{tabular}{c}
\includegraphics[width=200pt]{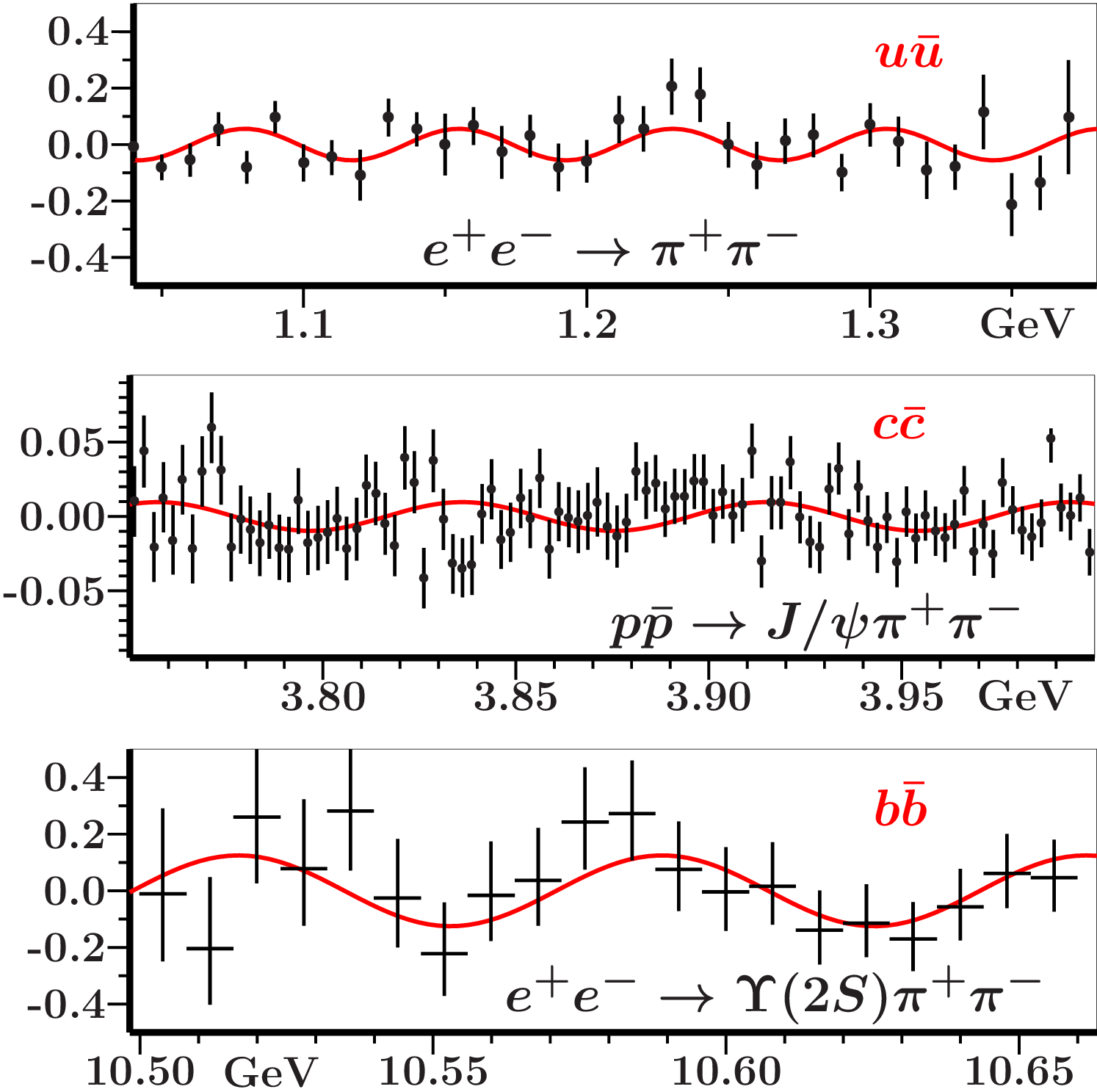}\\ [-10pt]
\end{tabular}
\end{center}
\caption{\small
Fits to the residual data, after subtraction of global fits to:
$e^{+}e^{-}\to\pi^{+}\pi^{-}$ data of the CMD-2 Collaboration
\cite{JETPL82p743}, with a period of 78$\pm$2 MeV
and an amplitude of $\approx$5\% ({\it top});
$p\bar{p}\to J/\psi\pi^{+}\pi^{-}$ data
of the CDF Collaboration \cite{PRL103p152001},
with a period of 79$\pm$5 MeV
and an amplitude of about 0.75\% ({\it middle});
$e^{+}e^{-}\to\Upsilon (2S)\pi^{+}\pi^{-}$ data
of the BaBar Collaboration \cite{PRD78p112002},
with a period of 73$\pm$3 MeV
and an amplitude of some 12.5\% ({\it bottom}).
}
\label{interference}
\end{figure}

In Ref.~\cite{PRD79p111501R}
notice was made of an apparent interference effect
around the $D_{s}^{\ast}\bar{D}_{s}^{\ast}$ threshold
in the invariant-mass distribution
of $e^{+}e^{-}\to J/\psi\pi^{+}\pi^{-}$ events,
which we observed in preliminary radiation data
of the BaBar Collaboration \cite{ARXIV08081543}.
The effect, with a periodicity of about 74 MeV,
could be due to interference
between the typical oscillation frequency of 190 MeV
of the $c\bar{c}$ pair,
as in the model of Refs.~\cite{PRD21p772,PRD27p1527},
and that of the gluon cloud.
Later, in Ref.~\cite{ARXIV10095191},
evidence was reported of small oscillations
in electron-positron and proton-antiproton annihilation data,
with a periodicity of 76$\pm$2 MeV, independent of the beam energy.
The latter observations are summarised in Fig.~\ref{interference}.

Amongst the various scenarios to explain the phenomenon
presented in Ref.~\cite{ARXIV10095191},
one was rather intriguing, namely the postulated existence of
gluonic oscillations, possibly surface oscillations,
with a frequency of about 38 MeV. These would then, upon interfering
with the universal quarkonium frequency $\omega =190$ MeV
\cite{PRD21p772,PRD27p1527},
lead to the observed oscillations.
However, here we will show that the phenomenon
is most likely to be associated with the
interquark exchange of a boson with a mass of about 38 MeV.
Moreover, from the fact that the observed oscillations
are more intense for bottomonium than for light quarks,
we assume that the coupling of this light boson to quarks
increases with the quark mass.
This seems to correspond well to the scalar particle
of the model of Ref.~\cite{NCA80p401},
and to the enigmatic mass parameter related to
the $^{3\!}P_{0}$ pair-creation mechanism \cite{PRD21p772}.

In Ref.~\cite{PRD78p112002}, the BaBar Collaboration
presented an analysis of data on
$e^{+}e^{-}$ $\to$ $\pi^{+}\pi^{-}
\Upsilon\left( 1,2\,{}^{3\!}S_{1}\right)$
$\to$ $\pi^{+}\pi^{-}\ell^{+}\ell^{-}$
($\ell =e$ and $\ell =\mu$),
with the aim to study hadronic transitions between
$b\bar{b}$ excitations
and the $\Upsilon\left( 1\,{}^{3\!}S_{1}\right)$
and $\Upsilon\left( 2\,{}^{3\!}S_{1}\right)$,
based on 347.5 fb$^{-1}$ of data
taken with the BaBar detector at the PEP-II storage rings.
The selection procedure for the data is well described by BaBar
in Refs.~\cite{PRD78p112002,PRL104p191801,ARXIV09100423}.

\begin{figure}[htpb]
\begin{center}
\begin{tabular}{c}
\includegraphics[width=240pt]{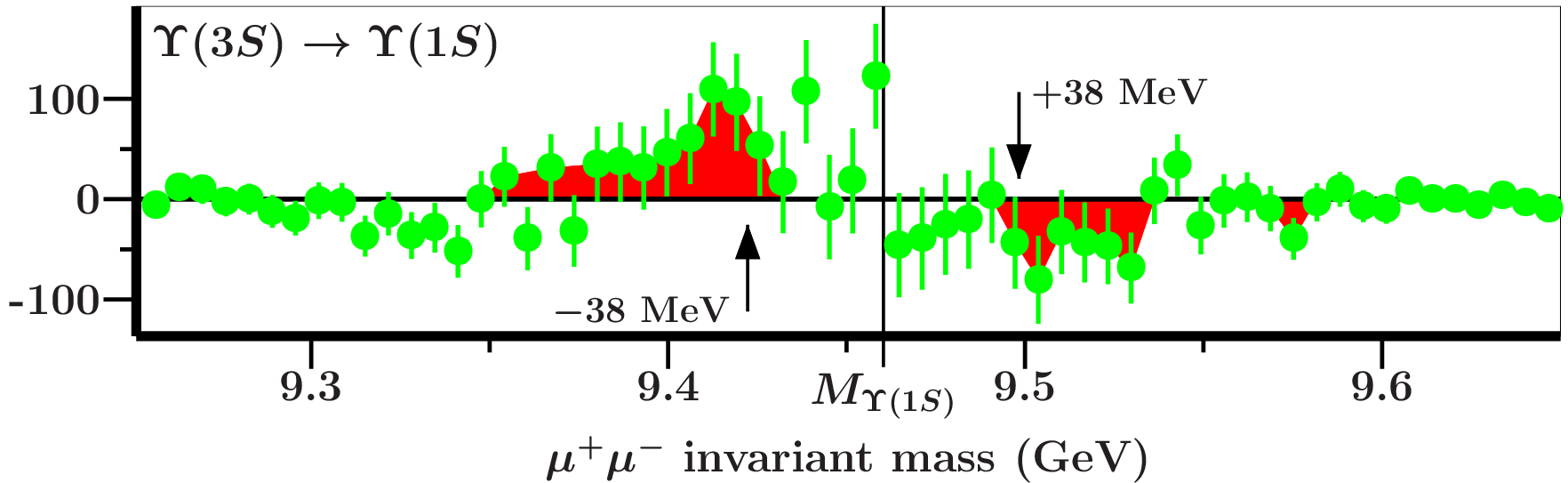}\\ [-10pt]
\end{tabular}
\end{center}
\caption{\small
Event distribution of the excess signal
taken from Ref.~\cite{ARXIV09100423},
in the invariant-$\mu^{+}\mu^{-}$-mass distribution
for the reaction
$\Upsilon\left( 2\,{}^{3\!}S_{1}\right)$ $\to$
$\pi^{+}\pi^{-}\Upsilon\left( 1\,{}^{3\!}S_{1}\right)$
$\to$ $\pi^{+}\pi^{-}\mu^{+}\mu^{-}$.
Statistical errors are shown by vertical bars.
The shaded areas (dark, red in online version)
are discussed in the text.
The vertical line indicates
$M_{\mu^{+}\mu^{-}}=M_{\Upsilon\left( 1\,{}^{3\!}S_{1}\right)}$.
}
\label{elisa}
\end{figure}
A particularly interesting study,
published by the BaBar Collaboration \cite{ARXIV09100423},
is the asymmetry with respect to a Gaussian distribution
for the reaction $e^{+}e^{-}$ $\to$
$\Upsilon\left( 2\,{}^{3\!}S_{1}\right)$ $\to$
$\pi^{+}\pi^{-}\Upsilon\left( 1\,{}^{3\!}S_{1}\right)$
$\to$ $\pi^{+}\pi^{-}\mu^{+}\mu^{-}$.
We depict the result in Fig.~\ref{elisa}.
We observe that, with respect to the Gaussian distribution,
there is an excess of data
for $M_{\mu^{+}\mu^{-}}$ below
the $\Upsilon\left( 1\,{}^{3\!}S_{1}\right)$ mass,
and a deficit of data
for $M_{\mu^{+}\mu^{-}}$ thereabove.
Moreover, the analysis in Ref.~\cite{ARXIV09100423}
took all known possible origins of asymmetry into account.
Consequently, what is left (see Fig.~\ref{elisa})
cannot be explained by known physics.
Furthermore, it states that the, here reported,
systematic uncertainties due to the differences
between data and simulation in the processes
$\Upsilon\left( 1\,{}^{3\!}S_{1}\right)$ $\to$ $\tau^{+}\tau^{-}$
and
$\Upsilon\left( 1\,{}^{3\!}S_{1}\right)$ $\to$ $\mu^{+}\mu^{-}$
cancel, at least in part, in their ratio. This
implies that a similar excess is found
in the
$\Upsilon\left( 1\,{}^{3\!}S_{1}\right)$ $\to$ $\tau^{+}\tau^{-}$
decay.

\begin{figure}[htpb]
\centering
\sidecaption
\includegraphics[width=230pt]{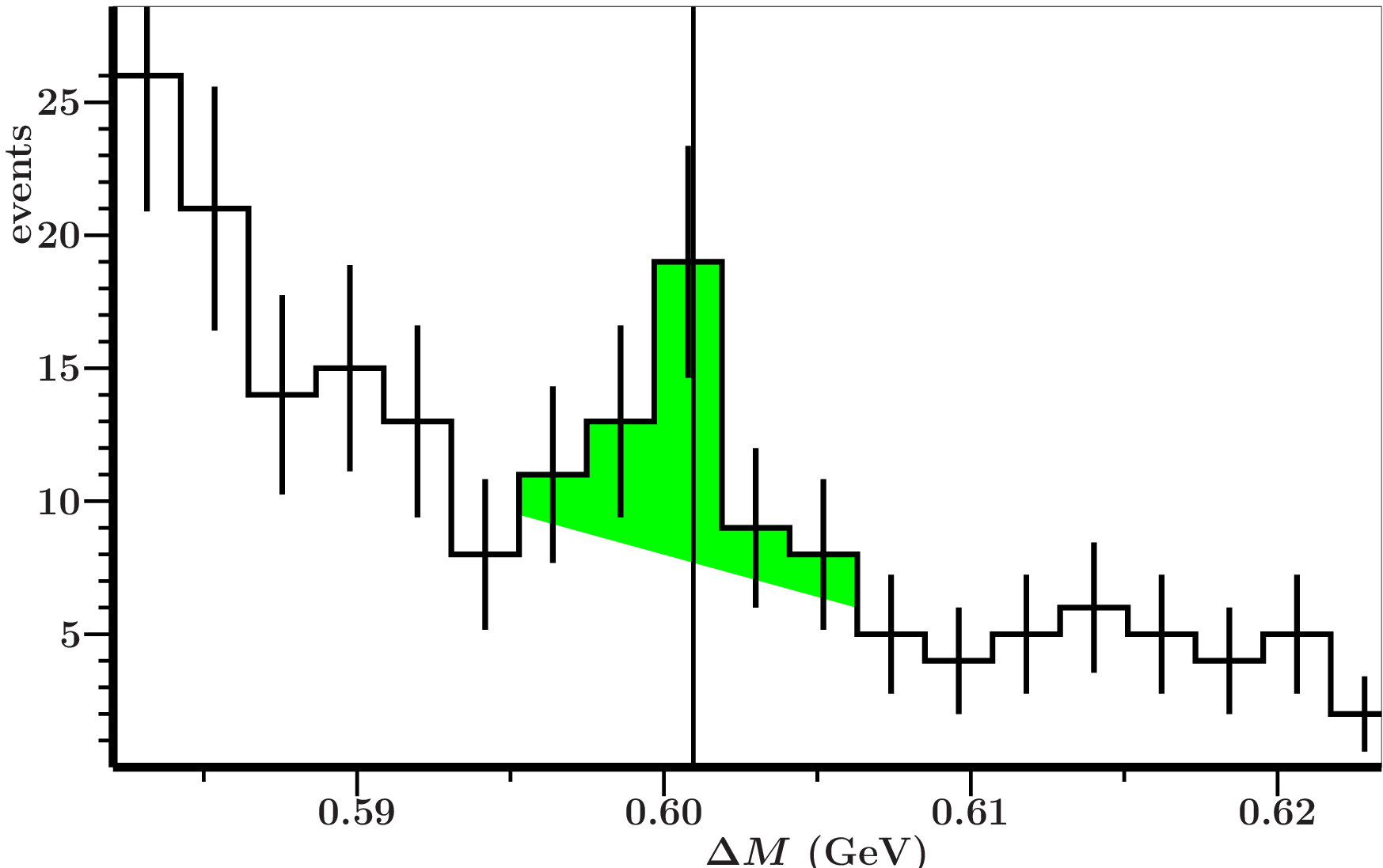}
\caption[]{\small
Possible sign of the $\Upsilon '\left( 2\,{}^{3\!}S_{1}\right)$ hybrid.
The vertical line indicates where
$M_{\Upsilon\left( 2\,{}^{3\!}S_{1}\right)}+38$ MeV comes out,
in terms of $\Delta M=M_{\pi^{+}\pi^{-}\mu^{+}\mu^{-}}-M_{\mu^{+}\mu^{-}}$.
Data are from Ref.~\cite{PRD78p112002}.
}
\label{hybrid}
\end{figure}

In order to explain the structures in the deficit signal,
we must assume that the $E(38)$ can be loosely bound
inside a $b\bar{b}$ state, giving rise to a kind of hybrid.
This was discussed to some detail in Ref.~\cite{ARXIV11021863}.
In Fig.~\ref{hybrid}, we show the event distribution
for the invariant mass $\Delta M$, which is defined
\cite{PRD78p112002} by $\Delta M=$
$M_{\pi^{+}\pi^{-}\mu^{+}\mu^{-}}-M_{\mu^{+}\mu^{-}}$,
where the latter mass is supposed
to be the $\Upsilon\left( 1\,{}^{3\!}S_{1}\right)$ mass.
Thus, a signal with the shape of a narrow Breit-Wigner resonance
seems to be visible on the slope of the
$\Upsilon\left( 2\,{}^{3\!}S_{1}\right)$
resonance, though with little more than 2$\sigma$ relevance.
Nevertheless, by coincidence or not, it comes out exactly in the
expected place, namely at
$M_{\Upsilon\left( 2\,{}^{3\!}S_{1}\right)}+38$ MeV.

\subsection{Diphoton mass distributions}
\label{gammagamma}

We do not expect the $E(38)\to\gamma\gamma$ mode
to be very large, since $E(38)$ couples to quarks proportionally to their
masses, as we concluded above in connection with the observed oscillations.
Moreover, diphoton data for the mass interval 10--100 MeV
are very rare and usually with low statistics.
\begin{figure}[htpb]
\centering
\sidecaption
\includegraphics[width=230pt]{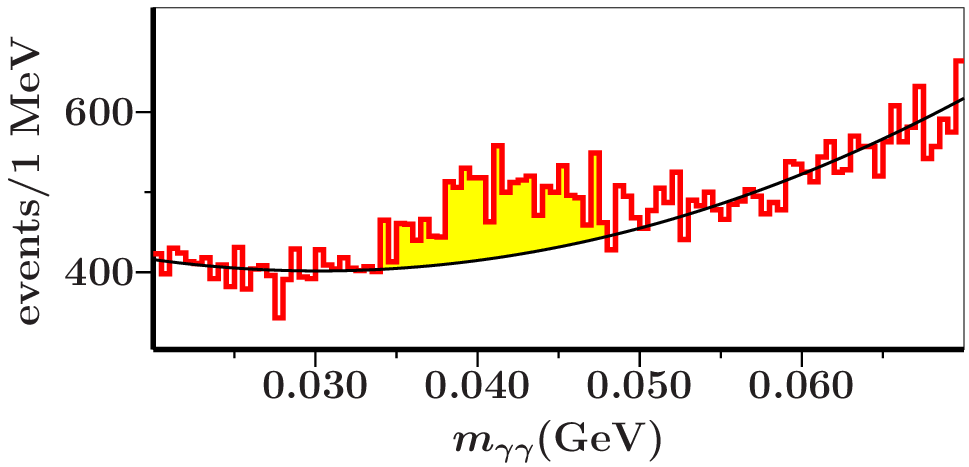}
\caption[]{\small
A modest signal in the $\gamma\gamma$ COMPASS \cite{ARXIV11090272} data
around 40 MeV.}
\label{bewaar}
\end{figure}

In Ref.~\cite{ARXIV11090272}, the COMPASS Collaboration
studied $\omega$ and $\phi$ vector-meson production in
$pp\to pp\omega /\phi$ data,
obtained at the two-stage magnetic COMPASS spectrometer
attached to the SPS accelerator facility at CERN.
For the indentification of the $\omega$ meson, the $\pi^{0}$
was reconstructed from two photons.
A detail of the thus obtained invariant-mass distribution
for $\gamma\gamma$ pairs is shown in Fig.~\ref{bewaar},
in which an enhancement at about 40 MeV can be observed.

These data seem to have enough statistics to support the existence of
a light boson with mass around 40 MeV.
However, in a more recent version \cite{ARXIV11090272} of their work,
the COMPASS Collaboration added the following remark
to the figure caption regarding the enhancement in
Fig.~\ref{compassgammagamma}:
\begin{quote}
\em
``The structures below the $\pi^{0}$ mass peak are artefacts
of low energetic photon reconstruction due to secondary interactions
in the detector material and to cuts in the reconstruction algorithm.
They should not be mistaken for any physical signal.''
\em
\end{quote}

Although it may of course be possible to obtain resonance-like structures
by artefacts in data collection, we are convinced this is not the case
for the signal in the 40 MeV region, because it coincides
surprisingly well with the other observations reported
in Refs.~\cite{ARXIV11021863,ARXIV12021739}.

Furthermore, it is clear that the light boson cannot be
an ordinary hadron emerging from a hadronic vertex,
unless at an extremely low rate.
Otherwise, it would have been observed long ago.
Ordinary hadronisation in high-energy collisions
gives rise to pions, kaons, and other hadrons that are stable with
respect to strong interactions.
These are processes in which quark-pair creation and gluonic jets dominate.
But on the other hand, judging from the amount of events in the
low-mass enhancement in the two-photon data,
which is about 10\% of the number of events in the $\eta$ signal,
it does seem to be produced with a reasonably large rate
in the COMPASS experiment.
Such a rate indicates that it most probably is a hadronic particle,
though with very peculiar properties that still have to be
understood.

Fortunately, the COMPASS Collaboration also presented
a Monte-Carlo (MC) simulation of the data in Ref.~\cite{ARXIV12042349}.
The authors of Ref.~\cite{ARXIV12042349} gave the following list
of mechanisms that may result in structures in their data.
\begin{itemize}
\item
Secondary $\pi^0$ mesons produced in the detector material
downstream of the target lead to diphoton masses which are
below the nominal $\pi^0$ mass when reconstructed
assuming a target vertex.
\item
Material concentrated in detector groups leads to peak-like structures.
\item
Secondary $e^+e^-$ pairs from photon conversion
in the spectrometer material lead to low-mass structures.
\item
Cuts applied in the reconstruction software lead to
additional structure in the low-mass range.
\end{itemize}
Most of those processes occur, of course, in the EM calorimeter
ECAL2, because it is further downstream than ECAL1,
and therefore picks up more contamination due to unwanted processes.

These artefacts are reproduced in the MC simulation
of Ref.~\cite{ARXIV12042349}
for the reactions under study, using a complete description
of the spectrometer material and employing
the same reconstruction software as for the real data analysis.

\subsection{Simulation of diphoton data}

\begin{figure}[htpb]
\centering
\sidecaption
\includegraphics[width=180pt]{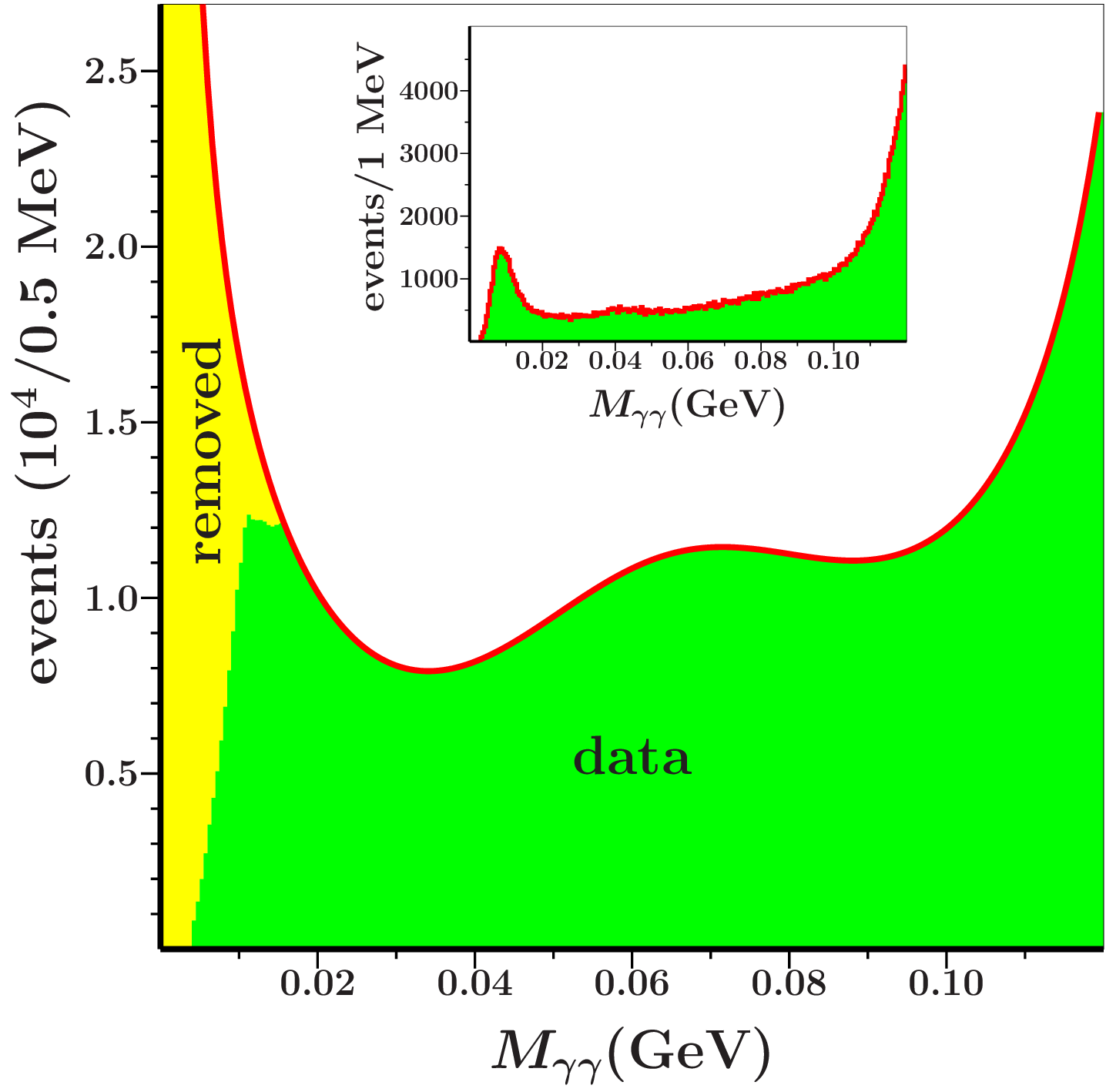}
\caption{\small
Invariant two-photon mass distributions below the nominal
$\pi^0$ mass. The solid curve in principle indicates the general aspect
of such a  distribution.
Data removed by the trigger system are represented
by the {\it removed} area.
So the {\it data} area in principle represents the final data.
In the inset, we display the diphoton mass distribution
for the $pp$ data of Ref.~\cite{ARXIV11090272}.
}
\label{ideal}
\end{figure}

In principle, the diphoton invariant-mass distribution
below the nominal $\pi^0$ mass has a structure as represented
by the solid curve in Fig.~\ref{ideal}.
However, the two electromagnetic calorimeters,
ECAL1 at about 6 meters from the target
and ECAL2 at about 30 meters downstream,
do not accept low-energy photons.
This results in the non-observation
of the enhancement at zero invariant diphoton mass.
Hence, at very low masses no events are observed,
since they are removed by the trigger system
of the EM calorimeters, as indicated by the {\it removed} area
in Fig.~\ref{ideal}.
It has the effect that at low mass a peak shows up in the data.

\begin{figure}[htbp]
\centering
\sidecaption
\includegraphics[width=180pt]{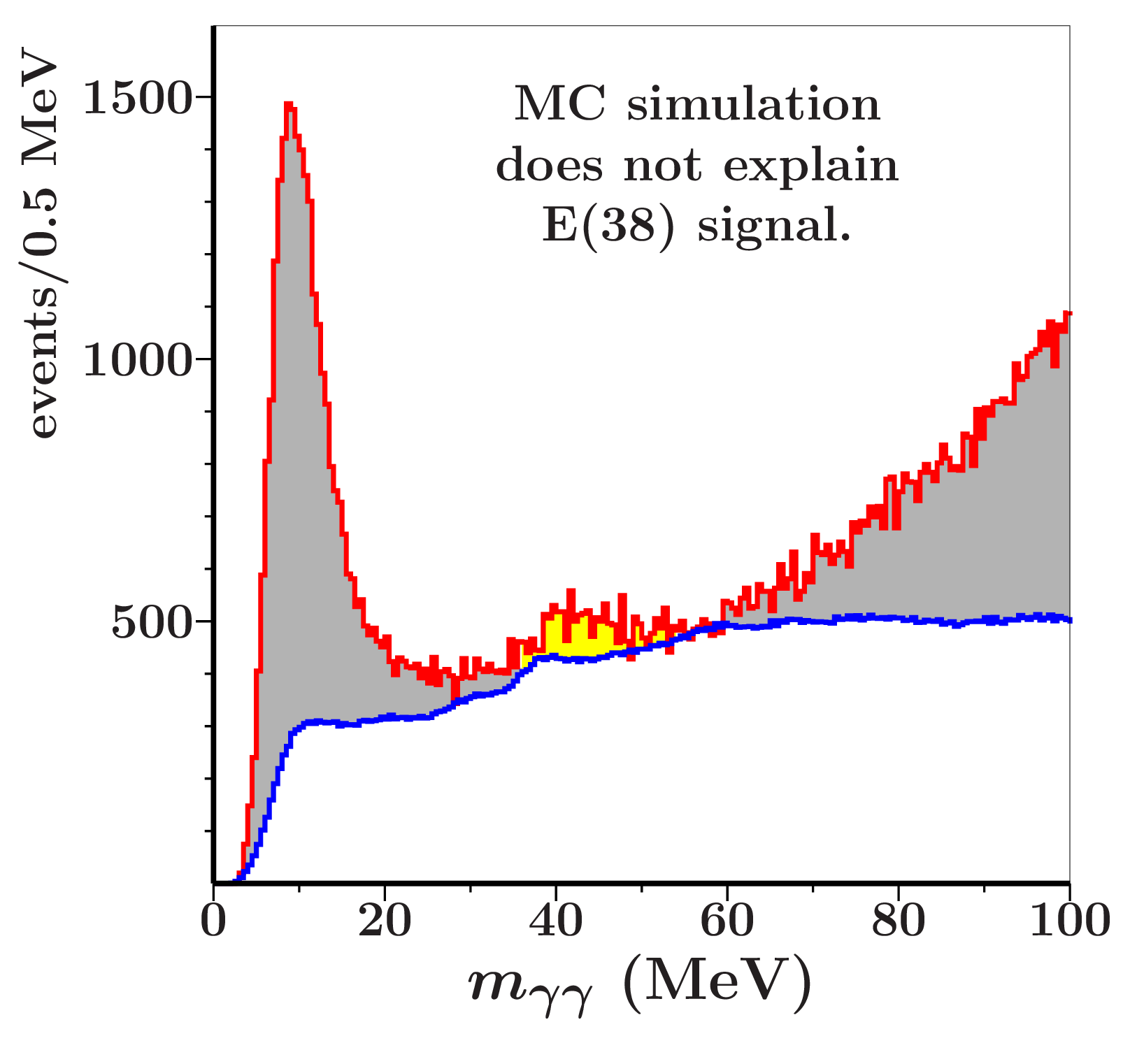}
\caption{\small
Diphoton mass distribution \cite{ARXIV11090272} (upper curve)
and Monte-Carlo simulation \cite{ARXIV12042349} (lower curve).
}
\label{ppjoinpi}
\end{figure}

Now, the data selection system does of course
also influence the aspect of the data for higher diphoton masses.
It may even result in some structures which resemble resonances.
Moreover, several physical processes in the experimental setup,
not related to the purpose of the experiment,
may result in further structures.
In Fig.~\ref{ppjoinpi} we depict the MC simulation
of Ref.~\cite{ARXIV12042349}.
We find that the experimental data
are not at all well described by the MC simulation
of Ref.~\cite{ARXIV12042349}.
The discrepancy between data and simulation at low diphoton masses
is rather serious and may indicate differences between the experimental
low-energy cuts and those of the simulation.
We observe from Fig.~\ref{ppjoinpi} that for some reason
the $\pi^{0}$ signal,
which in the experimental data sets out at about 50 MeV,
is not included in the simulations \cite{ARXIV12042349}.
From the discrepancy between the actual $\pi^0$ tail and
its simulation we may conclude that the spreading in the data
may not have been correctly taken into account in the simulation.
The latter observation would imply that the small effect
of the artefacts in the present simulation should be even smaller.
We may safely conclude that the MC simulation of
Ref.~\cite{ARXIV12042349} does not explain the signal
around 40 MeV \cite{ARXIV12043287}.

\subsection{High-statistics signal}

In Ref.~\cite{ARXIV11086191}, the COMPASS Collaboration
carried out a partial-wave analysis of $p\pi^{-}\to p\pi^{-}\eta '$
in order to extract the exotic $J^{PC}=1^{-+}$ $\pi^{-}\eta '$ $P$-wave
signal. The analysis consists of various intermediate steps.
The first step is to select $\pi^{0}\eta$ pairs,
possibly stemming from the decay of an $\eta'$-meson.
Either one of the two mesons $\pi^0,\eta$ or both may decay into pairs of
photons. Hence, as a byproduct of their analysis,
the COMPASS Collaboration produced an invariant-mass distribution
of photon pairs.
\begin{figure}[htpb]
\centering
\sidecaption
\includegraphics[width=180pt]{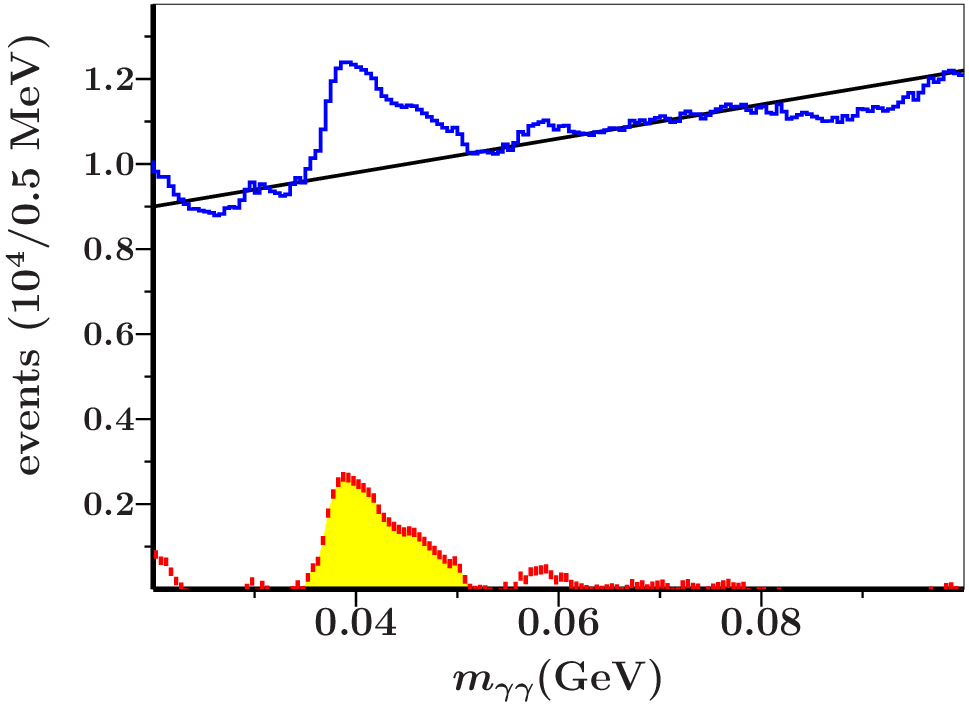}
\caption[]{\small
Top: a clear signal in the $\gamma\gamma$ COMPASS \cite{ARXIV11086191} data,
with maximum at about 39 MeV. Bottom: the $E(38)$ structure
that remains after background subtraction and contains about 46000 events.}
\label{compassgammagamma}
\end{figure}
In Fig.~\ref{compassgammagamma} we show a detail of
the invariant two-photon mass distribution of COMPASS.
These data seem to have enough statistics to substantiate the existence of
a light boson with mass around 40 MeV.

\section{Conclusions}

With respect to the hadronic sector of the Standard Model,
we have shown that high-statistics experiments are necessary
for serious data analysis,
in particular for the study of threshold enhancements.
Moreover, we have exhibited data that may indicate
the existence of a weak substructure.

The question whether there exists a (scalar) boson
with a mass of about 38 MeV does not depend exclusively
on the existence of a resonance-like structure
in the experimental data of Ref.~\cite{ARXIV11086191},
yet it is at present the clearest signal we have found in many
experiments. Diphoton data for the mass interval 10--100 MeV
are very rare and usually with low statistics.
Therefore, it is very important that the present issue be
settled, which requires a profound understanding
of all possible sources of artefacts.
In this respect, we welcome the efforts of our colleagues
at JINR (Dubna), who are reportedly finishing the analysis
of their $E(38)$ signal \cite{ARXIV12083829}.

With this latest piece of evidence of $E(38)$,
we conclude that it is now necessary to establish its mass
and other properties by further experiments.
We think that $E(38)$ is the light scalar Higgs-type boson
which was proposed in a model describing the unification
of electromagnetic and strong interactions \cite{NCA80p401}.
Finally, as $E(38)$ appears to couple to quarks
proportionally to their masses,
its coupling to the top quark is expected to be quite strong.

\section*{Acknowledgments}

This work was supported in part by the {\it Funda\c{c}\~{a}o para a
Ci\^{e}ncia e a Tecnologia} \/of the {\it Minist\'{e}rio da Ci\^{e}ncia,
Tecnologia e Ensino Superior} \/of Portugal, under contract
CERN/\-FP/ 123576/\-2011.
One of us (EvB) wishes to thank the organisers of the
{\it 3rd International Conference on New Frontiers in Physics}
for kindly inviting him to Kolymbari (Crete, Greece)
and for the hospitality at the {\it Conference Center
of the Orthodox Academy of Crete}.

\newcommand{\pubprt}[4]{#1 {\bf #2}, #3 (#4)}
\newcommand{\ertbid}[4]{[Erratum-ibid.~#1 {\bf #2}, #3 (#4)]}
\def\AIPCP{AIP Conf.\ Proc.}
\def\AdP{Annalen der Physik}
\def\AP{Ann.\ Phys.}
\def\CNPC{Chin.\ Phys.\ C}
\def\EPJC{Eur.\ Phys.\ J.\ C}
\def\EPL{Europhys.\ Lett.}
\def\IJTPGTNO{Int.\ J.\ Theor.\ Phys.\ Group Theor.\ Nonlin.\ Opt.}
\def\JETPL{JETP Lett.}
\def\LNP{Lect.\ Notes Phys.}
\def\NCA{Nuovo Cim.\ A}
\def\NPA{Nucl.\ Phys.\ A}
\def\NPB{Nucl.\ Phys.\ B}
\def\NPPS{Nucl.\ Phys.\ Proc.\ Suppl.}
\def\PL{Phys.\ Lett.}
\def\PLB{Phys.\ Lett.\ B}
\def\PRAMANA{Pramana}
\def\PRD{Phys.\ Rev.\ D}
\def\PRL{Phys.\ Rev.\ Lett.}
\def\ZPC{Z.\ Phys.\ C}

\end{document}